\documentclass[useAMS,usenatbib]{mn2e}
\usepackage{amsmath,amssymb,mathrsfs,graphicx,float,latexsym}
\usepackage{epstopdf,ragged2e}
\usepackage{bm,relsize} 
\usepackage{hyperref}
\usepackage[flushleft]{threeparttable}
\usepackage{subfigure}

\usepackage{color}
\newcommand{\Rs}{R_{\star}}
\newcommand{\Ms}{M_{\star}}
\newcommand{\Bc}{B_{\star}}

\newcommand{\bs}{\boldsymbol{\sigma}}
\newcommand{\PT}{\Phi^{\text{T}}}
\newcommand{\Mc}{M_{\text{comp}}}
\newcommand{\svom}{\sigma_{\text{vM}}}
\newcommand{\sm}{\sigma_{\text{max}}}
\newcommand{\Oo}{\Omega_{\text{orb}}}
\newcommand{\tC}{t_{\text{C}}}
\newcommand{\Bn}{B_{15}}
\newcommand{\Mn}{M_{1.4}}
\newcommand{\Rn}{R_{10}}

\newcommand{\tB}{t_{B}}
\newcommand{\Po}{\rho+p}
\newcommand{\Tn}{T_{90}}
\newcommand{\Tw}{T_{\text{wt}}}

\defcitealias{pap1}{SK20}
\defcitealias{Kuan21}{Paper I}

\begin{document}

\title[$g$-mode resonances as triggers for SGRB precursors]{General-relativistic treatment of tidal $g$-mode resonances in coalescing binaries of neutron stars. II. As triggers for precursor flares of short gamma-ray bursts}

\author[Hao-Jui Kuan, Arthur G.~Suvorov, and Kostas D.~Kokkotas]{Hao-Jui Kuan$^{1,2}$,\thanks{E-mail:hao-jui.kuan@uni-tuebingen.de} Arthur G.~Suvorov$^{1,3}$, and Kostas D.~Kokkotas$^{1}$\\
	$^1$Theoretical Astrophysics, Eberhard Karls University of T{\"u}bingen, T{\"u}bingen, D-72076, Germany\\
	$^2$Department of Physics, National Tsing Hua University, Hsinchu 300, Taiwan\\
	$^3$Manly Astrophysics, 15/41-42 East Esplanade, Manly, NSW 2095, Australia}

\date{Accepted XXX. Received YYY; in original form ZZZ}

\pagerange{\pageref{firstpage}--\pageref{lastpage}} \pubyear{2021}

\maketitle
\label{firstpage}

\begin{abstract}
 
\noindent{In some short gamma-ray bursts, precursor flares occurring $\sim$ seconds prior to the main episode have been observed. These flares may then be associated with the last few cycles of the inspiral when the orbital frequency is a few hundred Hz. During these final cycles, tidal forces can resonantly excite quasi-normal modes in the inspiralling stars, leading to a rapid increase in their amplitude. It has been shown that these modes can exert sufficiently strong strains onto the neutron star crust to instigate yieldings. Due to the typical frequencies of $g$-modes being $\sim 100\text{ Hz}$,  their resonances with the orbital frequency match the precursor timings and warrant further investigation. Adopting realistic equations of state and solving the general-relativistic pulsation equations, we study $g$-mode resonances in coalescing quasi-circular binaries, where we consider various stellar rotation rates, degrees of stratification, and magnetic field structures. We show that for some combination of stellar parameters, the resonantly excited $g_1$- and $g_2$-modes may lead to crustal failure and trigger precursor flares.}

\end{abstract}


\section{Introduction}

Short duration gamma-ray bursts (SGRBs), which are commonly defined as bursts having $90\%$ of their photon count detected with $\Tn\lesssim 2\text{ s}$ \citep{Kouveliotou93}, are thought to result from compact object mergers involving at least one neutron star (NS) \citep{Paczynski:1986px,Narayan:1992iy,Belczynski:2006br,Giacomazzo13}. Due to the complexity of the coalescence stages, i.e., inspiralling, merging, and ringdown, multi-stage measurements have been made for some SGRBs, including precursor flares, main episodes, and afterglows. Each delivers different information on NS physics, such as the equation of state (EOS) and the central engine of emissions from both the progenitors \citep{Giacomazzo13,Ascenzi19} and the remnants \citep{Lasky14,Sarin19,Suvorov20}. In particular, precursor flares have been observed for a few SGRBs with some of them likely occuring before the merger \citep{Tsang:2011ad}. Therefore, these fainter, though phenomenologically similar, flashes that precede the main episodes, offer an extra probe into the properties of progenitors on top of other means, such as gravitational-wave (GW) detections \citep{Hinderer10,Abbott18prl}.

It has been suggested that tidally-driven crust failures are responsible for precursors [\citealt{Tsang:2011ad,pap1} (SK20)]. Generally speaking, if the stress exerted on the stellar surface exceeds the maximum that the elastic crust can support, a yielding may be expected. We focus on NSNS binaries in this work, for which the external tidal field contributed by the companion star deforms the shape of the primary, inducing a quadrupole moment and certain crustal stress [see, e.g., Equations (64) and (65) of \cite{Ushomirsky00}]. In the final stage of inspiralling, the tidal field becomes tremendous and distorts the star to an extent that may lead to crustal failure [e.g., \cite{Owen05}], though the yielding resulting from this process can occur only within $\lesssim 10^2$ ms prior to the merger \citep{Kochanek92,Tsang:2011ad,Penner:2011br}. In light of the relative time difference of precursors to the main episodes, which ranges from a few hundred milliseconds to a few tens of seconds prior to the main episode [see, e.g., Table I in \cite{Kuan21} (henceforth Paper I) and references therein], the aforementioned equilibrium tidal effects are seemingly not capable of accommodating the observed precursors. On the other hand, stars are also deformed by dynamical tides, which are induced from the motion of matter, that can be decomposed into a sum of quasi-normal modes (QNMs). The tidal force drives QNMs at twice the orbital frequency \citep{zahn77}, which brings modes oscillating at the same frequency into resonance. The amplitudes of resonantly-excited modes increase rapidly as a consequence of their ability to efficiently absorb orbital energy over a resonance timescale. If a certain mode is driven so strongly that the resultant strain exceeds a critical value such that the crystalline structure of the crust can no longer respond linearly, a crustal failure may occur \citep{Horowitz09,Chugunov10,Baiko18}.

The crust failure liberates charged particles that are then accelerated by induction-generated electric fields to form ejecta. The outflow interacts with the surrounding medium, eventually leading to the conversion of magnetic energy flux into radiation
\citep{Blaes89,Thompson:1995gw,Spruit01}. Since precursors are observed to have a non-thermal spectrum \citep{Troja:2010zm,Zhong:2019shb} , (at least one of) the inspiralling stars should be highly magnetised ($B \gg 10^{13}$) so that the energy can be efficiently transported via Alfv{\'e}n waves \citep{Thompson:1995gw,Tsang:2011ad}.
Further credence is given to this scenario because magnetar birth rates \citep{Gullon:2015zca} coincide with the recent estimates on the proportion of SGRBs preceded by at least one precursor flare \citep{Troja:2010zm,Minaev:2018prq,cop20}. In view of these points, mode excitations in magnetars are worth exploring as they may be the central engine for these precursors (\citealt{Troja:2010zm,Tsang:2011ad}; \citetalias{pap1}).

In addition to magnetic fields, the stellar stratifications, and rotation, as well as EOS, adjust the inertial frame frequencies of QNMs, which in turn changes the timing of resonances. A search for realistic circumstances that connect with the observed precursors may thus shed light on the magnetic field structure, the rotation rate, and the EOS of progenitors [see, e.g., \cite{Neill21}]. It is important therefore that realistic models of crust yielding due to mode resonance be constructed, so that astrophysical information concerning NS structure can be extracted from precursor phenomenology. 
	
As introduced in \citetalias{Kuan21}, we extend previous frameworks (\citealt{Tsang:2011ad}; \citetalias{pap1}) utilized to study tidally-driven crustal yieldings as triggers of precursors in several aspects. Specifically, stellar QNMs are solved relativistically and the orbit evolution involves up to 3rd order post-Newtonian (PN) order effects including a 2.5PN flux scheme for gravitational back-reaction. Furthermore, mixed poloidal-toroidal magnetic fields together with rotational and stratification effects are also included numerically in our evolutions [see \citetalias{Kuan21} for more details]. The present article, together with \citetalias{Kuan21}, is devoted to a detailed evaluation of realistic scenario by cooperating all the aforementioned factors with the hope that they can eventually lead to predictions. 

Although not considered in the present article and \citetalias{Kuan21}, we note that it is commonly accepted that outer layers of a cold NS consist of a solid crust, which may lead to the quenching of modes at the crust-core interface [see, e.g., \cite{McDermott88,Levin01,Colaiuda11}]. In particular, the attenuation of the fluid motions due to $g$-modes within the crust implies a reduction in the crustal strain available during the resonance, and may thus weaken the relevance of $g$-mode resonances as the potential mechanism behind the precursors. The investigation of the influence of the crust entails a significant modification to the formalism in \citetalias{Kuan21}, and will be addressed elsewhere.
  
This work is organised as follows: In Section \ref{sec.II}, we illustrate how $g$-mode resonances may be associated to the properties of the precursor flares. In Section \ref{sec.III}, we compute the configuration of crustal strain by resonant modes and estimate the energy stored in the region where crust yields. The dependence of crustal strain on various stellar parameters is explored in Section \ref{sec.IV}, and some discussion is offered in Section \ref{sec.V}.

Unless stated otherwise, quantities are given in geometrical units with $c=G=1$. The greek letters refer to the four dimensional spacetime indices except $\alpha$, which denotes the quantum number of eigenmodes. The following abbreviations are adopted throughout: $\Bn=\Bc/(10^{15}G)$, $\Mn=\Ms/(1.4\text{ } M_{\odot}),\Rn=\Rs/(10\text{ km})$, and $E_{45}=E/(10^{45}\text{ erg})$.

\section{Precursor flares of short gamma-ray bursts} 
\label{sec.II}
GRBs show a bi-modal distribution in their durations, $T_{90}$, and are therefore often classified into two classes -- long ($T_{90}>2$) and short ($T_{90}<2$) \citep{Kouveliotou93}. Classifying a given event however is not trivial, because one should take the duration, redshift, other observations [e.g.~precursors, afterglows \citep{Nakar07}] and/or the possible limits of instruments [e.g.~duration of measurement in different energy bands \citep{Bromberg13}] into account [see the discussion in \cite{Berger14}]. Nonetheless, a simple but broadly used method to distinguish the short from the long is $\Tn\lesssim 2\text{ s}$ \citep{Kouveliotou93,Paciesas99,Jespersen20}.

Although rare, precursor flares are sometimes seen before SGRB. The identification of these precursors from the main episode depends sensitively on the definition of preemissions. Therefore, the proportion of SGRBs hosting precursor activities varies within literature. For instance, some authors require that a genuine precursor flare has to precede the main episode by more than $\Tn$ \citep{Troja:2010zm,Minaev17}, whereas some allow for arbitrarily short periods of time prior to the main burst for preemissions to be classified as precursor status \citep{Burlon08,Zhong:2019shb,cop20,Wang:2020vvr}. In Table \ref{tab:sgrbdata} we present relevant properties for the most statistically significant SGRB precursor candidates discussed in the above references. In the first column we show the associated SGRBs, and the second toward the penultimate ones are, respectively, the duration of the main bursts, the timing of precursor emissions prior to the main episodes (waiting time, $\Tw$), and the statistical significance. The final column  lists the inferred orbital frequency $\Omega_{\mathrm{orb}}$ by matching the time of the events with the binary evolution (Sec.~\ref{sec.II.A}), which indicates the frequency of the corresponding resonantly-excited mode (Sec.~\ref{sec.III}). We see that GRBs 071030, 090510b, 100717 and 130310 are temporally-separated, relative to the main burst, by at least couple of seconds ($\Tw\gtrsim2.5\text{ s}$), while others are prior to the main burst only within $\lesssim 1.85 \text{ s}$.

There are three events in \cite{Wang:2020vvr} having rather small or large $\Tw$, viz.~GRBs 100223110, 150922234 and 191221802. The first two precede the main episode by, respectively, $\gtrsim 80 \text{ ms}$ and 30 ms, and the waiting time for the latest is $\Tw\gtrsim 20 \text{ s}$. The closeness to the merger blurs the identification of the former two, i.e., these preemission may not proceed the merger since the formation timescale of the main emission is likely comparable or longer than $80 \text{ ms}$ (see Sec.~\ref{sec.II.A} for the discussion). On the other hand, the latter happens at a very early stage ($a\gtrsim200$ km), where the interaction between two stars in a binary, which is proportional to $a^{-3}$, is so weak that the mechanism behind this preemission may not be relevant to mutual interaction (unless the main burst was significantly delayed).

\begin{table*}
	\caption{Properties of SGRB precursor candidates as reported in \protect\cite{Wang:2020vvr,Minaev:2018prq,Zhong:2019shb,Troja:2010zm}. The associated orbital frequencies are determined by the time prior to the main burst, which is assumed to happen immediately after the merger (i.e., $\tB\approx\tC$). We assume an equal-mass binary that comprises stars with EOS SLy (see Sec.~\ref{sec.III.A}) and $M=1.27M_\odot=\Mc$ and $\Rs=11.78$ km [the same system used in Figure (4) of \protect\citetalias{Kuan21}]. The binary evolution is solved according to the numerical scheme in Fig.~3 of \protect\citetalias{Kuan21}, which involves up to 3PN terms in conservative orbital dynamics and 2.5 PN radiation-reaction. Tidal effects of $f$-modes are also taken into account. }
	\label{tab:sgrbdata}
	\begin{threeparttable}
	\begin{tabular}{ccccc}
	\hline
	\hline
	Precursor Event & Duration [$\Tn$ (s)] & Time prior to main burst [$T_{\text{wt}}$ (s)] & Significance ($\sigma$) & Orb. freq. [$\Oo$ (Hz)] \\
	\hline
	GRB 060502B & $\sim 0.09$ & $0.32$ & $6.1$ & 573.74 \\
	\hline
	GRB 071030 & $\lesssim 0.7$ & $2.5$ & $6.3$ & 283.89 \\
	\hline
	GRB 081216531 & $0.15^{+0.05}_{-0.03}$ & 0.53$^{+0.04}_{-0.05}$ & $>4.5$ & 484.27$^{-11.78}_{+16.50}$  \\
	\hline
	GRB 090510a \tnote{a}& $0.05 \pm 0.02$ & $0.45 \pm 0.05$ & $\lesssim 4.6$ & 511.78$^{-17.87}_{+20.68}$ \\
	GRB 090510b\tnote{a} & $\lesssim 0.4$ & $13$ & $5.2$ & 158.41 \\
	\hline
	GRB 100213A & $\sim 0.44$ & $0.68$ & $11.1$ & 445.00 \\
	\hline
	GRB 100717 & $0.3 \pm 0.05$ & $3.3$ & $12.8$ & 257.58 \\
	\hline
	GRB 100827455 & $0.11^{+0.05}_{-0.04}$ & $0.34\pm0.06$ & $>4.5$ & 562.24$^{-29.77}_{+37.54}$  \\
	\hline
	GRB 101208498 & $0.17^{+0.12}_{-0.08}$ & $1.17^{+0.10}_{-0.14}$ & $>4.5$ & 369.51$^{-10.31}_{+16.56}$ \\
	\hline
	GRB 111117510 & $0.18^{+0.05}_{-0.03}$ & $0.22^{+0.03}_{-0.06}$ & $>4.5$ & 649.61$^{-26.84}_{-71.34}$ \\
	\hline
	GRB 130310 & $0.9 \pm 0.32$ & $4.45 \pm 0.8$ & $10$ & 231.85${}^{-1.44}_{+1.29}$ \\
	\hline
	GRB 140209A & $\sim 0.45$ & $1.06$ & $13.9$ & 359.25 \\
	\hline
	GRB 141102536 & $0.06^{+0.10}_{-0.06}$ & $1.26^{+0.11}_{-0.15}$ & $>4.5$ & 360.18$^{-10.27}_{+16.09}$ \\
	\hline
	GRB 150604434 & $0.17{}^{+0.25}_{-0.01}$ & $0.64{}^{+0.02}_{-0.29}$ & $>4.5$ & 454.27$^{-4.72}_{+102.56}$ \\
	\hline
	GRB 160726A & $\sim 0.08$ & $0.39$ & $10.2$ & 537.00 \\
	\hline
	GRB 170802638 & $0.15^{+0.17}_{-0.11}$ & $1.85^{+0.14}_{-0.21}$ & $>4.5$ &  315.31$^{-7.9}_{+13.48}$\\
	\hline
	GRB 181126413 & $0.72^{+0.18}_{-0.27}$ & $0.85^{+0.40}_{-0.29}$ & $>4.5$ & 412.36$^{-51.19}_{+62.97}$ \\
	\hline
	\hline
	\end{tabular}
\medskip
\textbf{Notes:}
\begin{tablenotes}
	\item[a] GRB 090510a and GRB 090510b are not the official names of these two precursors. We label them by $a$ and $b$ to indicate the later and the earlier preemission episodes of GRB 090510.
\end{tablenotes}
\end{threeparttable}
\end{table*}

\subsection{Precursor Timing}\label{sec.II.A}

In reality, the main burst, occurring at $\tB$, will not be coincident with the coalescence at $\tC$, since the jet constituting the main burst has a finite formation timescale. Rather, the physical picture after the merger is complicated with several timescales participating in the SGRB mechanism, e.g., jet formation, jet break out, and GRB formation. In addition, each timescale varies with jet mechanism, making it almost impossible to make a conclusive statement about the separation between $\tB$ and $\tC$ [see Tab.~1 in \cite{Zhang19} for more details]. Though $\tB-\tC$ ranges from $0.01$ to $\lesssim$ 10 s, we assume that the burst occurs simultaneously with the merger, i.e.~$\tB\approx\tC$, with a caveat that the timing of precursor prior to the coalescence obtained under this assumption is actually the upper limit.

We consider a close NSNS binary system with constituent masses $\Ms$ and $M_\text{comp}$ for the primary and companion, respectively. The coalescence is defined to occur when the separation of binaries $a \lesssim 3q^{1/3}\Rs$ (\citealt{Lai93,Lai94,Ho99}; \citetalias{pap1}), where $\Rs$ is the radius of the primary and $q$ is the mass ratio $\Mc/M$ of the binary. The binary is evolved numerically until the point of coalesence defined above by including 3rd order post-Newtonian (PN) effects and GW back-reaction induced from the orbit in the 2.5 PN order and from excited modes\footnote{While $f$-modes are likely to get resonant before merging for rapidly rotating primaries \citepalias{pap1}, in this work we only slow rotation, thus $f$-mode resonances are absent. However, it has been shown that tidal effects of $f$-modes are important in binaries evolution \citep{Kokkotas:1995xe,Yang:2018bzx,Pratten20,Nijaid21} mainly due to their strong couplings with the tidal field. } [see \citetalias{Kuan21} in this series for details, see also \cite{Ogawaguchi96}].
The impacts of $p$- and $g$-modes on the binary evolution are, however, expected to be negligible as there is no resonance for the former and tidal couplings of the later is too small to affect binary evolution \citep{Shibata93}. In principle the $p$- and $g$-mode can couple to each other strongly due to their similar radial wavelength, resulting in so-called $p$-$g$ instability that may affect the binary evolution in a measurable way, e.g.,~heating up the star to $\gtrsim 10^{10}$ K and causing significant orbital phase errors \citep{Weinberg13,Essick16}. However, in the recent analysis of the gravitational wave event GW170817, the $p$-$g$ instability seems to either be suppressed to induce only slight phase shifts to the gravitational waveform or to be difficult to distinguish the effects from other intrinsic parameters of GW170817 \citep{Abbott19,Reyes20}. Therefore, we do not consider these effects on the evolution, and ignore the nonlinear tidal effects in the evolution equations [cf.~Sec.~3 of \citetalias{Kuan21}].

For a particular equal-mass binary ($q=1$) inspiralling on the equatorial plane\footnote{In close binaries, tidal interaction rapidly aligns the stellar spins with the orbital angular momentum \citep{Hut81,Zahn08}. Therefore, the inclination angle is expected to be approximately zero.} (i.e., the companion sits in the plane $\Theta=\pi/2$ with respect to the inertial frame of the primary throughout the evolution) with both stars obeying the SLy EOS (see Sec.~\ref{sec.III.A}), we determine the orbital frequencies at the moment precursors occur. In \citetalias{pap1}, a Newtonian scheme was used, i.e., by the Kepler formula,
\begin{align}\label{eq:kep}
	\Omega_{\text{Kep}} = \sqrt{\frac{(\Ms+\Mc)}{a^3}}.
\end{align}
Here, however, we use a PN scheme for orbital evolution and take the relativistic tidal effects into account. First, the coalescence is expedited, thus the orbital frequency at a certain time prior to merger is less; secondly, the mode eigenfrequencies, which are relevant for precursor timing, are shifted. As a consequence, the inferred (PN) orbital frequencies ( the last column of Tab.~\ref{tab:sgrbdata}) are found to be less than the inferred Keplerian orbital frequency in \citetalias{pap1} by $\lesssim10\%$ of the PN orbital frequencies. 

In addition, the frequencies of $f$-modes are $\gtrsim 2\text{ kHz}$ and the typical frequencies of $g_1$-modes are $\gtrsim 100\text{ Hz}$. Since the tidal force perturbs stars at a frequency which is twice the value of the orbital one (see Section \ref{sec.III.B} for details), the final column of Tab.~\ref{tab:sgrbdata} suggests, therefore, that the precursors are observed at the stage of inspiral prone to resonances of $g$-modes. 

\section{Resonant Shattering}\label{sec.III}

\subsection{Sellar models}\label{sec.III.A}

We consider a static, spherically symmetric spacetime whose line element reads
\begin{equation} \label{eq:statsphsym}
	ds^2 = - e^{2 \Phi(r)} dt^2 + e^{2 \lambda(r)} dr^2 + r^2( d\theta^{2}+\sin^{2}\theta^{2}d\phi^{2}) ,
\end{equation}
with $(t,r,\theta,\phi)$ being the usual Schwarzschild coordinates, and $\Phi$ and $\lambda$ being functions of $r$ only. The equations of motion for the star are then determined by the conservation laws $\nabla^{\mu} T_{\mu \nu}=0$, which require a specific EOS to complete the system of equations. We use the five EOS listed in \citetalias{Kuan21} to construct stellar models for they pass the constraints set by GW170817 \citep{Abbott18prl}, namely APR4 \citep{APR}, SLy \citep{SLy}, and three members of the WFF family\citep{WFF}. There is no \emph{a priori} reason why perturbations need to abide by the same EOS as the background. In particular, we allow for the perturbed density and pressure profiles to satisfy a different (barotropic) relation, i.e., the perturbation over the equilibrium can obey a different EOS \citep{Lockitch01}.

We consider non-isentropic perturbations, as relevant for $g$-modes, that are parameterized by introducing a free parameter $\delta$. Specifically, the adiabatic indices of the equilibrium $\gamma$,
\begin{align}
	\gamma=\frac{\Po}{p}\frac{dp}{d\rho},
\end{align}
 and that of the perturbation $\Gamma$ are related via
\begin{align}
	\Gamma=\gamma(1+\delta),
\end{align} 
where $\rho$ and $p$ are the density and the pressure profiles of the equilibrium. We note that, in principle one is able to extract the composition gradient from realistic EOS thus access $\Gamma$. Since we consider these EOS to be barotropic, i.e., $p=p(\rho)$,  the information of chemical composition is eliminated. Therefore, the artificially defined $\delta$ substantially describes \emph{non-adiabatic} perturbations \citepalias{Kuan21}.

 The stability of non-radially pulsating stars is determined by the Schwarzschild discriminant,
\begin{align}
	A = e^{-\lambda}\frac{dp}{dr}\frac{1}{p}\bigg( \frac{1}{\gamma}
	-\frac{1}{\Gamma} \bigg),
\end{align}
or equivalently the Brunt-Väisälä frequency,
\begin{align}
	N^2 = \tilde{g} A,
\end{align}
which is the characteristic frequency of the local fluid oscillations [see, e.g., \cite{Kokkotas99}]. Here $\tilde{g}$ is the local acceleration of gravity. When $N^2$ is positive, the fluid element oscillates around its equilibrium position, while the fluid is locally unstable where $N^2$ is negative \citep{Detweiler73}. Accordingly, we assume $\delta$ to be positive as otherwise the perturbation is unstable.

We imbue the star with an equilibrium magnetic field that is constructed so that (i) the field is dipolar, (ii) the field matches to a force-free dipole outside of the star ($r>\Rs$), and (iii) there is no surface current [Sec.~4 in \citetalias{Kuan21}].
The magnetic field that fulfils the above conditions and matches to the Schwarzschild exterior is uniquely found to be
\begin{equation} \label{eq:magf}
	B^{\mu} = \Bc \left( 0, \frac {e^{-\lambda}} {r^2 \sin\theta} \frac{\partial\psi}{\partial\theta}, 
	- \frac {e^{-\lambda}} {r^2 \sin\theta}  \frac{\partial\psi}{\partial r}, 
	- \frac {\zeta(\psi) \psi e^{-\Phi}} {r^2 \sin^2\theta}  \right),
\end{equation}
where $\psi(r,\theta) = (a_{1} r^2 + a_{2} r^4 + a_{3} r^6)\sin^2\theta$, and
\begin{equation}
	\zeta(\psi) \psi = - \left[ \frac{E^{p} \left( 1 - \Lambda \right)} {E^{t} \Lambda} \right]^{1/2}
	\frac{\left( \psi - \psi_{c} \right)^2}{\Rs^3}
	\label{eq:torratio}
\end{equation}
when $\psi \geq \psi_{c}$, and $\zeta$ is zero otherwise. In Eq.~\eqref{eq:torratio}, $\psi_c$ is the value of $\psi$ of the last closed field line interior the star.
Constants $a_1-a_3$ can be found in Eq.~(48) of \citetalias{Kuan21}, and we avoid repeating them here.
In Eq.~\eqref{eq:magf}, $0<\Lambda\le1$ parametrizes the ratio between the poloidal $E^{p}$ and toroidal energies $E^{t}$.
By the argument of the minimal energy among configurations with a constant magnetic helicity \citep{Bekenstein87}, the stable magnetic field has the toroidal-to-poloidal ratio $10^{-3}\lesssim\Lambda\lesssim0.3$ \citep{Akgun13,Herbrik17}.

Though not addressed in the present article, cold NSs may possess superfluid components that coexist with the crust lattice due to neutron drip, and/or locate at the core rendered by the exotic matter, e.g., hyperons and deconfined quarks [see, e.g., \cite{Andersson21}]. Superfluidity alters the structure of NSs in several aspects such as the induction equation that governs the perturbations in magnetic field \citep{Lander13}, and the $g$-mode spectrum \citep{Yu17}. Resulting GWs from binaries that contains at least one NS with superfluid may therefore be influenced \citep{suv21}.

\subsection{Tidal Resonance}\label{sec.III.B}
The tidal field sourced by the companion excites the quasi-normal modes of the primary, where leading-order terms are the $l=m=2$ components of tidal potential, viz.(\citealt{zahn77,Willems03}; \citetalias{pap1}), 
\begin{align}
	\PT = -\frac{\Mc}{8r}\bigg( \frac{r}{a} \bigg)^{3}P^{2}_{2}(\cos\theta)e^{2i\phi}e^{i\eta t},
\end{align}
with $\eta=2\Omega_{\text{orb}}$ being the forcing frequency. The free mode structure is determined by the EOS, and comprises $p$-, $f$-, $g$-, and $w$-modes. Each mode is labeled by the ensemble of quantum numbers $\alpha=(nlm)$ for overtone number $n$, and spherical harmonic indices $l$ and $m$. Time-dependent displacements of matter elements relative to their equilibrium places are composed of QNMs, and can be expressed as
\begin{align}
	\xi=\sum_{\alpha}q_{\alpha}(t)\xi_{\alpha},
\end{align}
where $q_{\alpha}$ are the QNM amplitudes, and the eigenfunctions $\xi_{\alpha}$ can be decomposed into radial ($\xi^r$) and poloidal ($\xi^h$) harmonics as \citep{Chandrasekhar64,Thorne67,Detweiler85}
\begin{align}
	\xi_{\alpha} = \bigg(\xi^{r}_{nl},\xi^{h}_{nl}\frac{\partial}{\partial\theta},\xi^{h}_{nl}\frac{1}{\sin^2\theta}\frac{\partial}{\partial\phi}\bigg)Y_{lm}.
\end{align}

A perturbing force $\delta F^{\mu}$ introduces a shift into the (inertial frame) mode frequencies according to 
\begin{align}\label{eq:freqpert}
	\delta\omega_{\alpha}=  \frac{1}{2\omega_{\alpha}}
	\frac{\mathlarger{\int_{\text{primary}}\delta F_{\mu}\overline{\xi}^{\mu} \sqrt{-g}d^{3}x}}
	{\mathlarger{\int_{\text{primary}}(\rho+p)e^{-2\Phi}\xi^{\mu}\overline{\xi}_{\mu}\sqrt{-g}d^{3}x}},
\end{align}
where $\xi_{\alpha}$ is the displacement of the QNM, $\omega_{\alpha}$ is the unperturbed (i.e., free mode) frequency, and the integral is taken over the volume of the primary. The overhead bar denotes complex conjugation.
In \citetalias{Kuan21}, we introduced the effects of magnetic fields, stellar rotations, and tidal fields on the free QNM spectrum, which are summarised as follows [see Sec.~5 of \citetalias{Kuan21} for the relevant equations and the full derivation]:
\begin{enumerate}
	\item Perturbations of the magnetic field, $\delta B^{\mu}$, by a certain QNM [Eq.~(57) of \citetalias{Kuan21}] generates a Lorentz force $\delta F^{\mu}_{B}$ [Eq.~(53) of \citetalias{Kuan21}] on the equilibrium, resulting in the frequency modulation
	\begin{align}
		\delta\omega^B_{\alpha}
		=&\frac{(\Ms\Rs^{2})^{-1}}{8\pi\omega_{\alpha}}\int_{\text{primary}} \sqrt{-g} d^{3}x \bigg[
		-\omega_{\alpha}^{2}B^{2}\xi^{\mu}\overline{\xi}_{\mu}e^{-2\Phi} \nonumber\\
		&+2B_{\mu}\delta B^{\mu}\overline{\xi}^{r}\Phi' 
		-\overline{\xi}_{\mu}\nabla_{\nu}\bigg(B^{\mu}\delta B^{\nu}+B^{\nu}\delta B^{\mu}\bigg) \nonumber\\
		&+\overline{\xi}^{\nu}\nabla_{\nu}(B_{\mu}\delta B^{\mu})
		\bigg].
		\label{eq:modmag}
	\end{align}
	Here $B^{\mu}$ is the equilibrium magnetic field and the overhead bar denotes complex conjugation.
	\item The leading-order tidal force,
	\begin{align}
		\delta F^{T}_{\mu} =\frac{\Mc}{a^{3}}( \Po )\nabla_{\mu}(\PT),
	\end{align}
	gives rise to
	\begin{align}
		\delta\omega^{T}_{\alpha}=\frac{Q_{n2}\Mc}{2\omega_{\alpha}a^{3}},
		\label{eq:modtid}
	\end{align}
	where the tidal overlap ($Q_{n2}$) is a complex, \emph{dimensionless} measure of the tidal coupling strength of the mode and is defined by \citep{Press77}
	\begin{align}
		Q_{n2} &=\frac{1}{\Ms\Rs^{2}}\int_{\text{primary}} d^{3}xe^{\Phi+\lambda} (\Po)  
		\overline{\xi}_{n22}^{\mu} \nabla_{\mu}(r^{2}  Y_{22}) r^{2}.
		\label{eq:overlap}
	\end{align}
	\item Rotation gives rise to a centrifugal force [Eq.~(69) in \citetalias{Kuan21}],
	resulting in a rotating frame frequency modulation of $m\Omega C_{nl}$ thus a inertial frame frequency modulation as
	\begin{align}\label{eq:modrot}
		\delta\omega^{R}_{\alpha} = -m\Omega (1-C_{nl}),
	\end{align}
	with
	\begin{align}
		C_{nl}= \frac{1}{\Ms\Rs^2}\int_{\text{primary}} (\Po)e^{\Phi+\lambda}r^{2l-2}
		 (\xi_{r}\overline{\xi}_{h} + \overline{\xi}_{r}\xi_{h} + \xi_{h}\overline{\xi}_{h}) dr.
	\end{align}
\end{enumerate}
In Eq.~\eqref{eq:modrot}, $\Omega=\nu/2\pi$ is the angular frequency of the stellar spin.

Including all the aforementioned perturbing forces, the resonance of a particular mode whose total (inertial frame) frequency is
\begin{align}
	\omega_{\text{tot}}=\omega_{\alpha}+\delta\omega_{\alpha}^{B}+\delta\omega_{\alpha}^{T}+\delta\omega_{\alpha}^{R},
\end{align}
occurs when the orbital frequency satisfies
\begin{align}\label{eq:resdef}
	|1-2\Oo/\omega_{\text{tot}}|\lesssim\epsilon,
\end{align} 
for some small parameter $\epsilon$ \citep{Lai94}.
In our case of $g$-mode resonances, we set, in the numerical point of view (as illustrated in \citetalias{Kuan21}, see Fig.~4 therein), the small parameter $\epsilon$ to be
\begin{align}\label{eq:resnbh}
	\epsilon=10\sqrt{  \frac{2\pi }{\Omega_{\text{orb}}}  \frac{|\dot{a}|}{a}     },
\end{align} 
where $\dot{a}$ is the temporal derivative of separation $a$. Nonetheless, we note that the physical resonance determined by \eqref{eq:resdef} is not necessarily captured by $\epsilon$ so defined in Eq.~\eqref{eq:resnbh} if the resonant frequency differs much from the range that we focus on in this article. The resonance duration $t_{\text{res}}$ is obtained as the time separation between the onset and the offset of resonance, viz.~the length of the time interval over which \eqref{eq:resdef} holds.

\begin{figure}
	\includegraphics[scale=0.45]{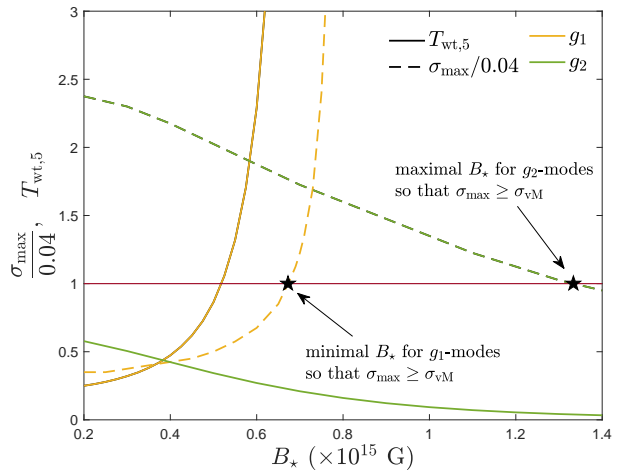}  
	\caption{Time prior to the main burst, which is assumed to coincide with the coalescence, $t$ (green and yellow solid lines) and maximal strain $\sm$ (dashed lines) as functions of $\Bc$. The black stars mark the minimal and the maximal value of $\Bc$, such that the von Mises criterion is met for $g_1$- and $g_2$-modes, respectively. We consider a binary with $q=1$ and the non-rotating primary having EOS SLy and $M=1.27M_{\odot}$. Here $T_{\text{wt},5}=T_{\text{wt}}/$(5 s).}
	\label{fig:diag}
\end{figure}

As a practical application of the resonant shattering scenario to observations, we consider a particular primary, within an equal-mass binary, with\footnote{The stratification $\delta =0.01$ we used to match the data of precursors in Tab.~\ref{tab:results} is higher than the typical value taken for NSs, which is $\delta=0.005$ \citep{Reisenegger:2008yk,Xu:2017hqo}. This degree of stratification may still be sensible for the resonances of high order modes along the inspiral rapidly absorb tidal energy. Besides, the dissipation of mode energy via GW is extremely inefficient for $g$-mode \citep{Finn87,McDermott83}, which ranges from tens to thousands of years for $g$-modes in this work. Therefore, the energy absorbed by high order $g$-modes will retain in the star until final merger with negligible dissipation.} $\delta=0.01$. The considered primary has a free $g_1$-mode resonance prior to the coalescence by $\sim 1 \text{ s}$, while the resonance of its free $g_2$-mode occurs at $\sim 3.2\text{ s}$ before the merger. For the stable range of $\Lambda$, the magnetic frequency modification is negative for the $g_1$-mode and is positive for the $g_2$-mode.
Setting $\Lambda=0.01$, we show $T_{\text{wt}}$ and $\sm$ for $g_1$- and $g_2$-modes as functions of $\Bc$ in Fig.~\ref{fig:diag}. We see that, when $\Bc$ approaches some certain values, both $\Tw$ and $\sm$ become dramatically larger for $g_1$-modes, as a consequence of the neutral frequency ($\omega_{\text{tot}}\to 0$) that triggers instability [cf.~Fig.~7 in \citetalias{pap1}]. Additionally, to account for those precursors occur within $1$ s prior to the merger, we vary $\Bc$ to match the resonant time of $g_2$-modes temporally with the aforementioned precursors. 

In Tab.~\ref{tab:results}, we show in the second column the characteristic strength of the magnetic field $\Bc$ such that the orbital frequency starts sweeping through the resonance interval defined by modified mode frequency, i.e., $|1-2\Oo/\omega_{\text{tot}}|\lesssim\epsilon$, at the moment the corresponding precursor occurs. The third towards the final columns are, respectively, the resonance duration $t_{\text{res}}$, the waiting time $T_{\text{wt}}$, the energy restored in the (crustal) region where crust yields (see Sec.~\ref{sec.IV.B}), and the orbital frequency inferred by the resonant time [We note that here $g_2$-mode resonances has been included, which was ignored in Tab.~\ref{tab:sgrbdata}]. With the same $\Bc$ as GRB 090510a, we find that for GRB 090510b one requires a stellar spin of $\nu=68.62 \text{ Hz}$ so that the inertial frame frequency is reduced, else it is impossible to match the mode frequency with its waiting time of $\Tw = 13 \text{ s}$. However, this implies an unphysically steep spin-down between GRBs 090510b and 090510a, i.e., $\Delta\nu=68.62$ in less than 13 s. The tension can likely be alleviated if we also consider a rotation for GRB 090510a, then we find a $\Bc$ such that the rotation rates responsible for GRBs 090510a and 090510b are not so different. In any case, we analysis GRB 090510b by using the same $\Bc$ as GRB 090510a in Tab.~\ref{tab:results}. 

\begin{figure}
	\includegraphics[scale=0.45]{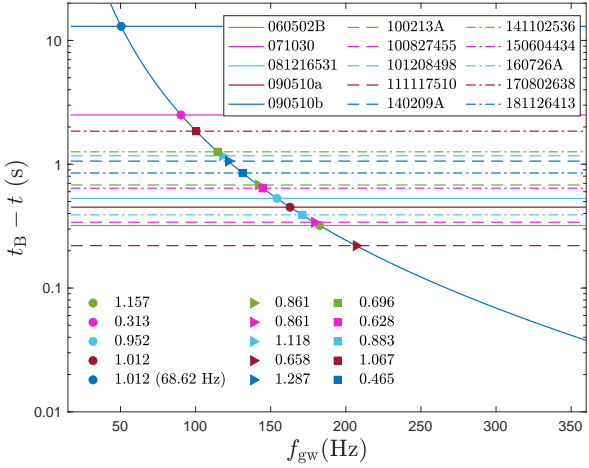}  
	\caption{Time prior to the main burst, which is assumed to coincide with the coalescence, as a function of gravitational wave frequency $f_{\text{gw}}=\Omega_{\mathrm{orb}}/\pi$. The solid line represents the evolution with tidal effect of $f$-mode but not $g_1$-mode for the non-rotating and non-magnetized star with EOS SLy and $M=1.27M_{\odot}$. The time of precursors reported in Tab.~\ref{tab:sgrbdata} are plotted as horizontal dashed lines. Markers are labeled by the characteristic magnetic field strengths $\Bc$ given in the unit of $\Bn$ for which the resonance frequencies of $g_2$-modes $\omega_{g}=2\pi f_{\text{gw}}$ match the precursor events. The number in the parenthesis is rotation rate for GRB 090510b.}
	\label{fig:SGRBs}
\end{figure}

In Fig.~\ref{fig:SGRBs}, we plot the precursors in Tab.~\ref{tab:results} labeled by $\Bc$ in the second column, and the overplotted curve represents the orbital evolution with only tidal effects of the $f$-mode. The label of GRB 090510b includes the rotation rate mentioned above. We see that the inferred orbital frequencies involving the resonances of $g_{2}$-modes [coloured symbols] are almost the same as the values predicted when only $f$-mode effects are considered [blue curve, the final column of Tab.~\ref{tab:sgrbdata}], reflecting the fact that the $g_2$-mode resonances barely affect the orbital evolution.

In addition, the resonances of the $g_1$-mode may precede the merger by more than 10 s for the toroidal-to-poloidal ratio in the range for a stable magnetic field, i.e.~$10^{-3}\lesssim\Lambda\lesssim0.3$ \citep{Akgun13,Herbrik17}. The same stands even when stellar rotation is considered since rotation decreases the frequencies of $l=2=m$ $g$-modes. Therefore, instead of appealing to stellar rotation to account for GRB 090510b, one may use the resonance of $g_1$-mode to account for GRB 090510, viz.~the two events could be accommodated by a $g_1$ and a $g_2$ excitations, respectively (see Sec.~\ref{sec.V.B} for details).


\begin{table*}
	\centering
	\caption{The relative quantities of the resonances matching temporally with listed precursors in Tab.~\ref{tab:sgrbdata}.We assume the resonances of $g_{2}$-modes for an equal mass binary, whose constituents obey the SLy EOS and have $\Ms=1.27M_{\odot}$. We set $\Lambda=0.01$, $\nu=0$ Hz, and $\delta=0.01$. GRBs 100717 and 130310 are not included since they are not suitable for the $g_{2}$-mode of this star.}
	\begin{threeparttable}
	\begin{tabular}{ccccccc}
			\hline
			\hline
			Precursor Event & $\Bc$ ($\Bn$)  & $t_{\text{res}}$ (s)  & $T_{\text{wt}}$(s)
			& $\sm$ & Fracture Energy ($E_{45}\cdot$s) & Orb. freq. [$\Oo$ (Hz)]\\ 
			\hline
			GRB 060502B & 1.157 &  0.140 & 0.32 & 0.047 & 1.60 & 573.23 \\ 
			\hline
			GRB 071030 & 0.313  &  0.286 & 2.50 & 0.091 & 1.13 & 283.49 \\ 
			\hline
			GRB 081216531 & 0.952 &  0.140 & 0.53 & 0.056 & 1.76 & 484.15  \\ 
			\hline
			GRB 090510a & 1.012  &  0.133 & 0.45 & 0.053 & 2.24 & 511.64  \\ 
			GRB 090510b & 1.012\tnote{a}  & 0.428 & 13.00 & 0.151 &  10.96 (+1.58) \tnote{b} & 158.41 \\ 
			\hline
			GRB 100213A & 0.861  & 0.152 & 0.68 & 0.061 & 1.70 & 445.02 \\ 
			\hline
			GRB 100827455 & 1.118  & 0.122 & 0.34 & 0.048 & 1.73 & 562.59 \\ 
			\hline
			GRB 101208498 & 0.658  & 0.180 & 1.17 & 0.072 & 1.63 & 369.74    \\ 
			\hline
			GRB 111117510 & 1.287  & 0.106 & 0.22 & 0.042 & 0.36 & 649.86 \\ 
			\hline
			GRB 140209A & 0.696  &  0.175 & 1.06 & 0.070 & 2.25 & 382.51 \\  
			\hline
			GRB 141102536 & 0.628 & 0.184  & 1.26 & 0.074 & 1.57 & 360.17 \\ 
			\hline
			GRB 150604434 & 0.883  &  0.148 & 0.64 & 0.059 & 1.96 & 454.32 \\
			\hline
			GRB 160726A & 1.067  &  0.127 & 0.39 & 0.051 & 1.91 & 537.55 \\ 
			\hline
			GRB 170802638 & 0.465 & 0.211  & 1.85 & 0.083  & 1.20 & 315.16 \\ 
			\hline
			GRB 181126413 & 0.779 & 0.162  & 0.85 & 0.065 & 1.70 & 412.62 \\ 
			\hline
			\hline
	\end{tabular}
\medskip
\textbf{Notes:}
\begin{tablenotes}
	\item[a] There is no $\Bc$ that can make the resonance happen at 13 s prior to the main burst; instead, for this event, we use the same $\Bc$ as GRB 090510a and vary the rotation frequency. Precursor time matches the resonant time as $\nu=68.62$ Hz.
	\item[b] The number in the parentheses is the rotational energy.
\end{tablenotes}
\end{threeparttable}
\label{tab:results}
\end{table*}

\section{Energetics} \label{sec.IV}
During the resonance, the mode amplitude increases rapidly, stretching the crust more strongly over time. Here the crust is defined to be the region ranging from $0.9\Rs$ to the stellar surface ($\gtrsim 1 \text{ km}$). The strain due to a QNM\footnote{The factor $2$ difference compared to the usual definition of the strain, i.e.~$\sigma \equiv \sqrt{ (\bs_{\mu\nu}  \overline{\bs}^{\mu\nu})/2}$ [see, e.g. \cite{Suvorov19}], results from the dual mode to $\xi_{\alpha}$ that has eigenfrequency $-\overline{\omega}_{\alpha}$ \citep{Andersson97}.},
\begin{equation} \label{eq:vonmises}
	\sigma_{\alpha} \equiv \sqrt{2 [q_{\alpha}(t)\bs_{\mu\nu}]  [\overline{q}_{\alpha}(t)  \overline{\bs}^{\mu\nu}] },
\end{equation}
is proportional to mode amplitude, where the general-relativistic strain tensor is defined as \citep{Carter72,Carter75,Xu01}
\begin{align} \label{eq:straindefn}
	\bs_{\mu\nu}
	= \frac {1} {2} \left( \partial_{\mu} \xi_{\nu} + \partial_{\nu} \xi_{\mu} 
	+ \delta g_{\mu\nu} \right) - \Gamma^{\sigma}{}_{\mu\nu}\xi_{\sigma}.
\end{align}
Denoting the maximal value of the crustal strain induced by a QNM when its amplitude reaches the peak during resonance as $\sm$, we assume the crust fails if the von Mises criterion, coming from classical elasticity theory \citep{ll70}, is met, i.e., when $\sm$ exceeds some maximal breaking strain $\svom$ that the crust can sustain.

\subsection{Breaking Strain}\label{sec.IV.A}
The critical value $\svom$ is hard to determine, and, in principle, it may depend on the duration of stress (or the timescale of the mechanism that generates the stress), density, temperature, and composition of the crust, and so on~\citep{Chugunov10}. In recent molecular dynamics simulations, by adopting Zhurkov's model for breaking mechanism, \cite{Chugunov10} found a universal expression for $\svom$ (see their Eq.~6). There are several combinations of density and temperature having been studied in literature, e.g., the crust with the density of $10^{13}$ g/cm$^3$ and $T=0.1$MeV$\approx10^{9} \text{K}$ corresponds to $\svom\approx 0.1$ \citep{Horowitz09}, while $\svom\gtrsim 0.11$ for a density of $10^{14}$ g/cm$^3$ and $T\gtrsim 10^{8}$ K \citep{Hoffman12}. 

\begin{figure*}
	\subfigure{	\includegraphics[scale=0.4]{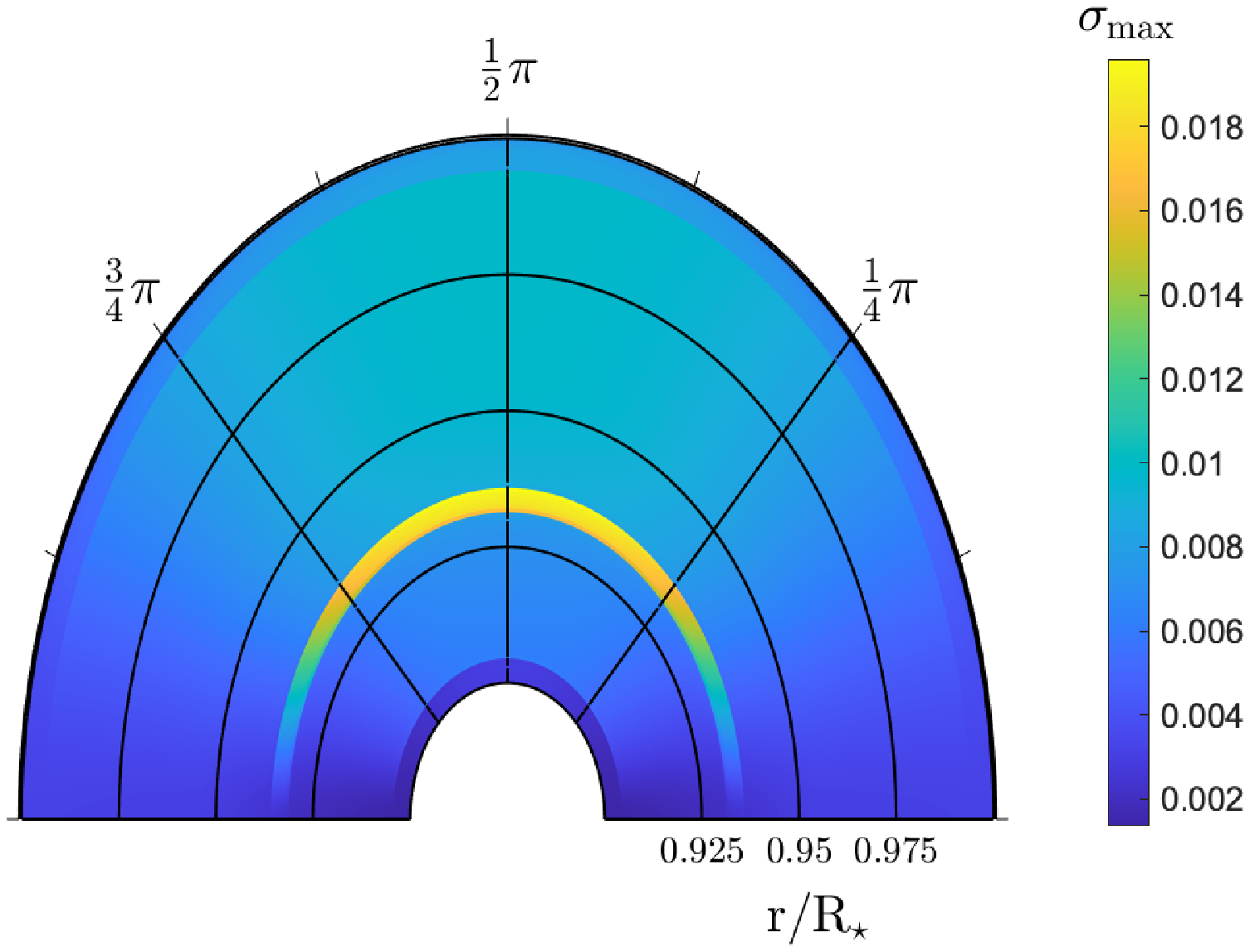}  }
	\subfigure{	\includegraphics[scale=0.4]{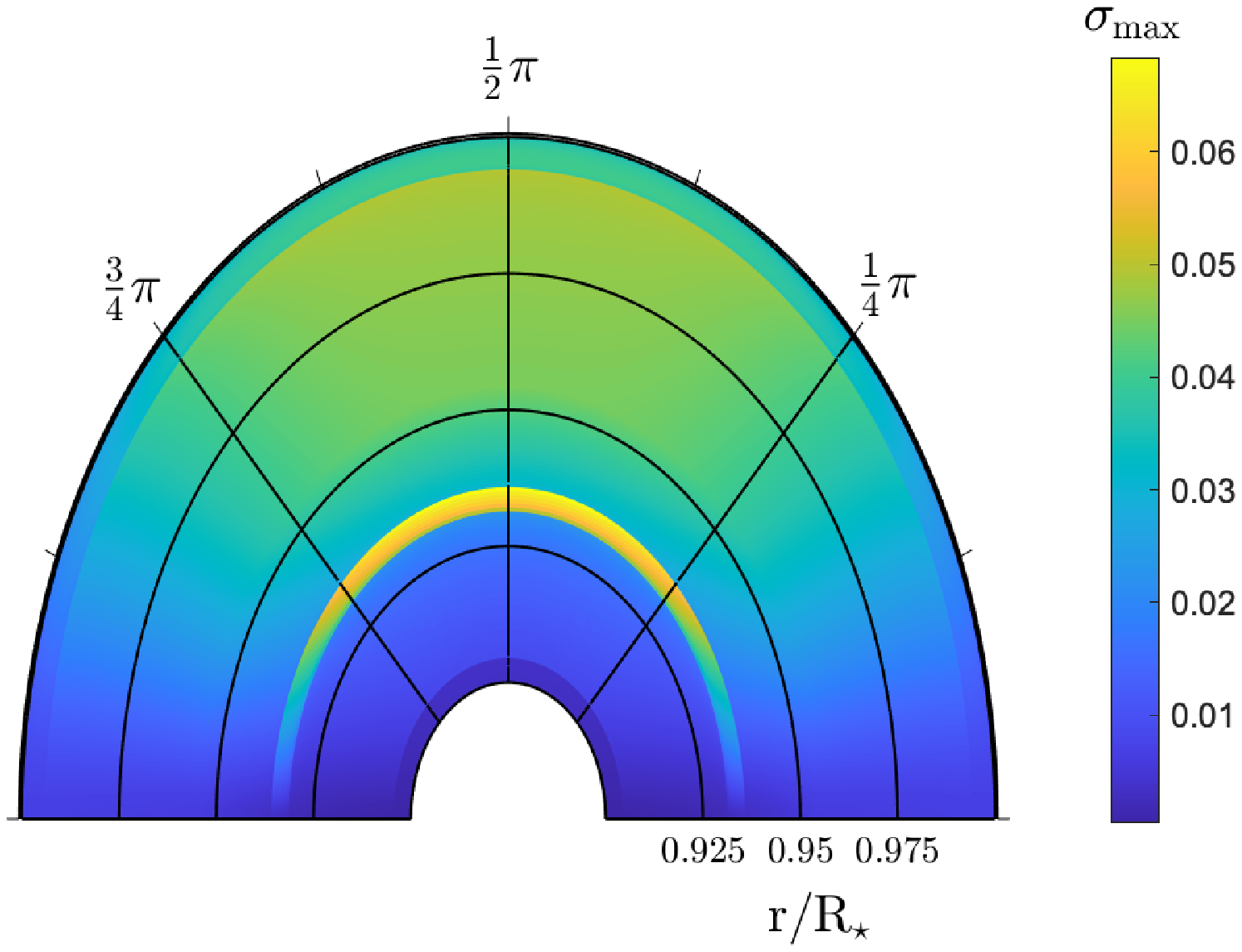}  }
	\caption{Configuration of crustal strain $\sigma$ by the $g_1$-mode (left panel) and the $g_2$-mode (right panel) of the non-magnetized star with EOS SLy having $\Ms=1.27M_{\odot}$ (the one used in simulating the orbital evolution in the final column of Tab.~\ref{tab:sgrbdata}) at the offset of resonance. We fix $\delta=0.005$ and adopt a log-linear grid to shrink the core region for illustration purposes.}
	\label{fig:strain_config}
\end{figure*}

In addition, \cite{Baiko18} follow a semi-analytical approach to calculate $\svom$ for low temperature stars [see also \cite{Baiko17}]. They found that, assuming the absence of the pasta phases, $\svom\sim 0.04$ which is density independent. In this work, we adopt $\svom\sim0.04$ as in \citetalias{pap1}, while we note that if $\svom\sim0.1$ had been adopted, crustal failure would entail a larger mode amplitude. In Fig.~\ref{fig:strain_config} we show the distribution of crustal strain generated by the $g_{1}$- and $g_{2}$-modes of the primary as a member of an equal-mass binary at the peak of resonance with some fixed stellar parameters. Both show that the region under the  relatively strong strain is narrow. Regions that fracture are restricted to the equatorial regions ($0.25\pi\lesssim\theta\lesssim0.75\pi$), indicating the crack is more likely to happen at these areas.

\subsection{Energy Release}\label{sec.IV.B}
To see if the precursor flares could fit in the context of SGRB precursors, it necessitates an estimation of the amount of energy potentially released due to crustal fracturing. 

Assuming that the onset of the resonance is at $t=0$ (this assumption is introduced for convenience and is valid only in this section), the liberated energy during the resonant shattering \citep{Lander19,Gourgouliatos21},
\begin{align}
	\int dt E_{\text{quake}} =\int_{0}^{t_{\text{res}}}dt \int_{V_{\text{crack}}(t)}\sqrt{-g}d^{3}x U(t,\mathbf{x}),
\end{align}
is obtained by temporally integrating the energy stored in the cracking area over the resonant duration $[0,t_{\text{res}}]$, where $U(t,\mathbf{x})$ is the energy density (see below).
In reality the cracking region, defined by 
\begin{align}
	V_{\text{crack}}(t)=\{ p\; |\; \sigma(p) \ge \svom, \; p\text{ is a point in the crust} \},
\end{align}
and the energy density, $U(t,\mathbf{x})$, are time-dependent. However, we approximate the energy released during a resonant timescale by integrating the energy density at the \emph{onset} of resonance over the cracking area at the \emph{offset} of resonance \citepalias{pap1}, namely
\begin{align}\label{eq:availerg}
	\int dt E_{\text{quake}} \approx t_{\text{res}} \times\int_{V_{\text{crack}}(t_{\text{res}})}\sqrt{-g}d^{3}x U(0,\mathbf{x}).
\end{align}

The available energy density includes the kinetic energy denisty of oscillation modes, $U_{\text{kin}}$ \citep{Thorne:1969}, the rotational energy density, $U_{\text{rot}}$ \citep{Hartle:1970,Morrison04}, the magnetic energy density, $U_{\text{mag}}$ \citep{ciolf}, and the tidal energy density, $U_{\text{tid}}$. The expressions for each of the contributions are, respectively, given by
\begin{subequations}
	\begin{align}
		&U_{\text{kin}} = \frac{1}{2} (\rho+p)e^{-2\Phi}  
		\frac{\partial\xi_{i}}{\partial t}\frac{\partial\overline{\xi}^{i}}{\partial t}, \\
		&U_{\text{rot}} = \frac{1}{2}\Omega^{2} r^{2}\sin^{2}\theta(\rho+p)e^{-2\Phi},\\
		&U_{\text{mag}} = \frac{1}{8\pi} e^{-\Phi} B^{\mu}B_{\mu},
	\end{align}
and
	\begin{align}
		U_{\text{tid}} = \PT\delta\rho, \hspace{2.6cm}
	\end{align}
\end{subequations}
where we reduce $U_{\text{rot}}$ to the uniform rotation case and the frame-dragging is not taken into consideration \citep{Belvedere14}.

 For those resonances explored in Tab.~\ref{tab:results}, the expected fracture energies are listed in the final column. We find that the kinetic energies of resonantly excited modes and the tidal energy contribute insignificantly ($\lesssim 10\%$) to the energy released as described by Eq.~\eqref{eq:availerg} unless the resonance onsets in the final stages of inspiral, $a\lesssim 6 \Rs$, in agreement with the findings of \citetalias{pap1}.

\section{Exploring the parameter space}\label{sec.V}

The duration and timing of mode resonances are influenced by various parameters, including the mass of the primary $\Ms$ and the companion $\Mc$ (or the mass ratio $q$ between them), stratification $\delta$, rotation frequency $\nu$, characteristic magnetic strength $\Bc$, the poloidal-to-toroidal strength $\Lambda$ and EOS.
This section is devoted to a detailed investigation of mode resonances over a multidemensional parameter space spanned by these parameters. 
In Sec.~\ref{sec.IV.A} we investigate the impact of $q$ on the maximal strain $\sm$ under fixed stratification strength $\delta$ and magnetic field. In Sec.~\ref{sec.IV.B} we assume equal-mass binaries to explore how other parameters affect $\sm$.

\subsection{Unequal-Mass Binaries} \label{sec.V.A}

We assume the same EOS for both of them, as that is the assumption adopted by \cite{Abbott18prl}. The magnetic field is considered to be purely poloidal ($\Lambda=1$), for which a field strength of a few $10^{15}\text{ G}$ is needed to shift the frequencies of $g_1$-modes by a noticeable amount while a few $10^{14}\text{ G}$ can already shift the frequencies of $g_2$-modes considerably [cf.~Fig.~(7) of \citetalias{Kuan21}]. Consequently, we set $\Bc=10^{15}\text{ G}$ in this section.
In addition, we fix $\delta=0.005$ to evaluate the maximal crustal strains of the primary by $g_{1}$- and $g_{2}$-modes. Restricting the masses of both components within the range\footnote{The considered range for $(\Ms,\Mc)$ covers a wide part of the parameter space compared to the NSs that have been observed [mostly from pulsar observation, cf.~Fig.~2 of \cite{Ozel:2016oaf}, see also Fig.~28 of \cite{Lorimer08}], which ranges from $\sim1M_{\odot}$ to $\sim 2M_{\odot}$ with a few outliers. However, we also consider stars with $M<1 M_{\odot}$ for completeness.} $(0.4,2.2)M_{\odot}$, we find for aforementioned EOS that despite increasing with $q$, $\sm$ depends only slightly on $q$ even in extreme cases. For $g_1$-modes, the difference in $\sm$ among binaries with a fixed $\Ms$ is $\lesssim0.005$, while the difference is $\lesssim0.01$ for $g_2$-modes.  For the pure poloidal magnetic field considered here, only those extreme cases of binaries, whose primaries have either $\Ms\gtrsim2M_{\odot}$ (for SLy,WFF1, and WFF3) or $\Ms\lesssim0.9M_{\odot}$ (for WFF2-3 and APR4), $\sm$ by $g_{1}$- or $g_{2}$-modes can achieve the von Mises threshold. 

While for a particular primary, $\sm$ depends only slightly on the companion (i.e., insensitive to $q$), the hosting-binary tends to have relatively small symmetric mass ratio,
\begin{align}\label{eq:symratio}
	q_{\text{sym}} =\frac{\Ms\Mc}{(\Ms+\Mc)^2}=\frac{q}{(1+q)^2},
\end{align} 
since the von Mises criterion is met for the primary with either large or small mass. Assuming a skewed normal distribution of the mass of NS in a NSNS binary, \cite{Kiziltan:2013oja} estimates the proportion of NSNS binaries having a component with mass out of the range $(\sim1.1,\sim1.55)M_{\odot}$ is less than $5\%$, while \cite{Ozel:2016oaf} assumes a normal distribution instead, the result is almost the same. In addition, the authors of the former reference find that less than $0.64\%$ for the mass lying out of $(\sim1,\sim1.7)M_{\odot}$. The rareness of precursor-hosting SGRBs, e.g.~$\sim0.5\%$ in \emph{Swift} data \citep{cop20} or $\sim 3\%$ in BATSE data \citep{Koshut:1995}, is compatible with the above estimation. Although the above point is certainly not conclusive, it does hint that a NSNS binary with relatively small symmetric mass ratio may be tied to precursor activity.

\begin{figure*}
	\includegraphics[scale=0.43]{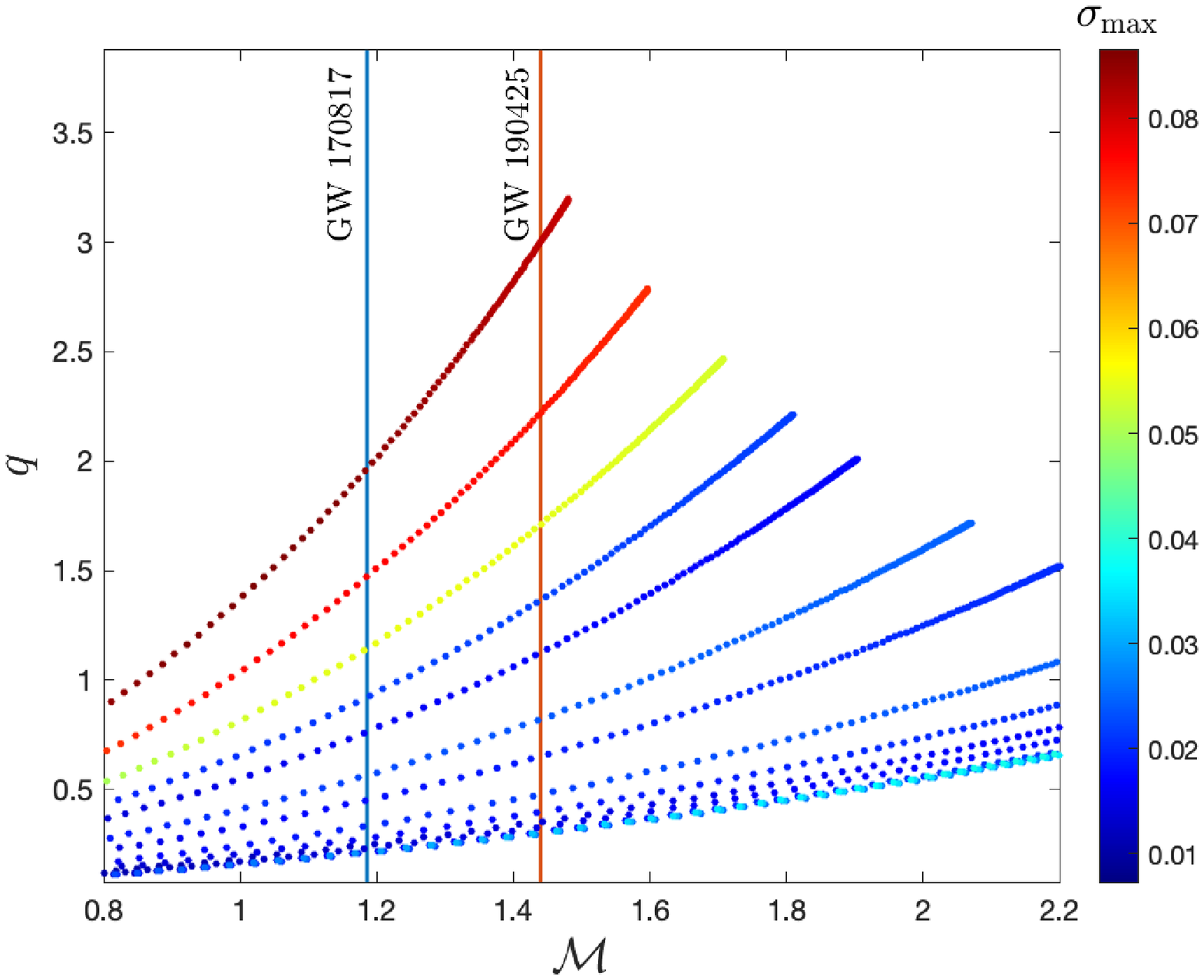}  
	\includegraphics[scale=0.43]{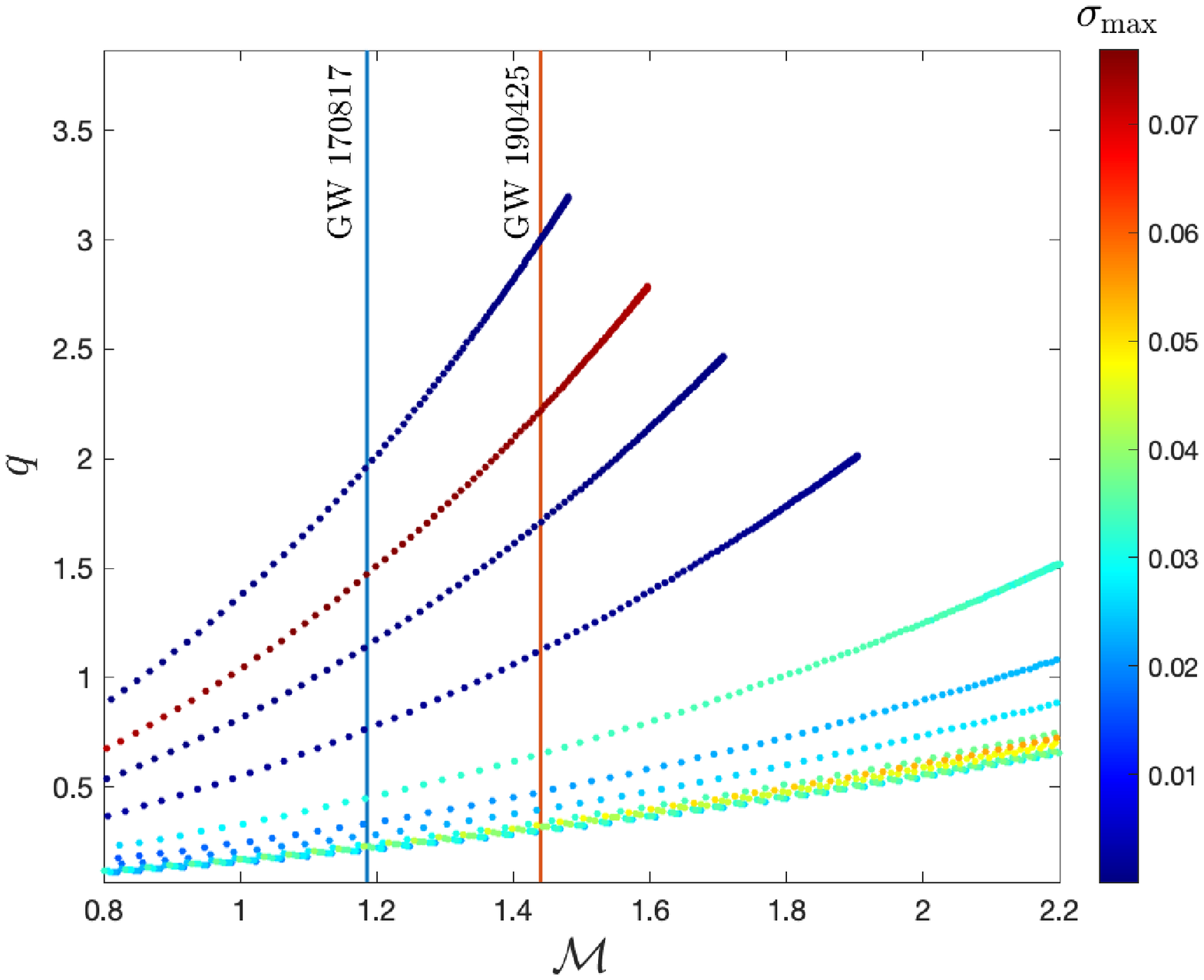}  
	\caption{Maximal strain $\sm$ by $g_1$-modes (left panel) and $g_2$-modes (right panel) available during a resonant timescale for systems of EOS WFF1 with several chirps masses $\mathcal{M}$ and mass ratios $q$. The blue vertical line shows the chirp mass of the progenitor of GW 170817, while the red similarly corresponds to GW 190425.
	}
	\label{fig:chirp-q-wff1}
\end{figure*}

In Fig.~\ref{fig:chirp-q-wff1} we plot the maximal strain $\sm$ available during the resonances of $g_1$- (left panels) and $g_2$-modes (right panels) for EOS WFF1, respectively, with a variety of  chirp masses \citep{Cutler94},
\begin{align}
	\mathcal{M} = \frac{(\Ms\Mc)^{3/5}}{(\Ms+\Mc)^{1/5}}=\Ms\frac{q^{3/5}}{(1+q)^{1/5}},
\end{align}
and mass ratios $q$. We see that for a fixed $\mathcal{M}$ , the von Mises criterion is met ($\sm\gtrsim\svom$) by $g_1$ and $g_2$ excitations for small or large $q$. For instance, if a binary with the WFF1 EOS has a chirp mass at the similar level of GW 170817 ($\mathcal{M}=1.186 M_{\odot}$), the resonances of $g_1$- and $g_2$-modes can generate $\sm\gtrsim\svom$ with a mass ratio $q\gtrsim 1.14$ and $q\lesssim 0.45$, respectively. In addition, the points with almost the same colour represent binaries with the same primary, which indicates that, for a given primary, $\sm$ depends mildly on $q$.

\begin{figure}
	\includegraphics[scale=0.45]{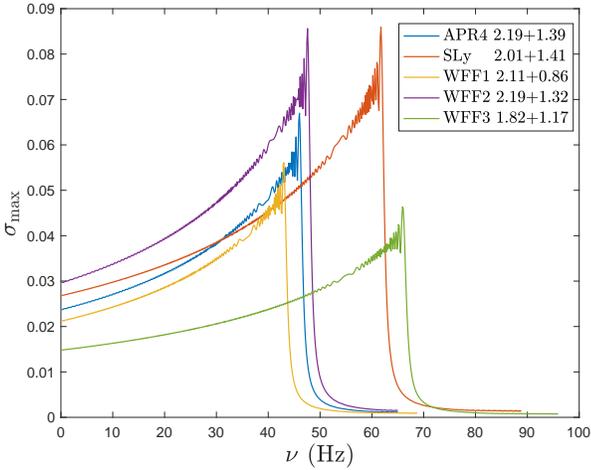}  
	\caption{Maximal crustal strain $\sm$ induced by $g_1$-modes as functions of rotating rate $\nu$. Labels of systems comprise the EOS that is obeyed by the primary and the companion, and the masses of these two components (in units of $M_{\odot}$).}
	\label{fig:sig_spin}
\end{figure}

The region in the parameter space over which crustal failure may occur will be expanded, viz.~more systems $(\Ms,\Mc)$ are likely to host a crack,  if stars rotate moderately or the magnetic field has strong enough toroidal component in that mode frequencies are shifted downward resulting in longer resonances. In Fig.~\ref{fig:sig_spin} we show $\sm$ as a function of $\nu$. We can see that these systems are capable of producing $\sm\gtrsim\svom$ at some certain range of $\nu$, e.g., when $31.33 \text{ Hz}\lesssim\nu\lesssim 46.54\text{ Hz}$ for the binary with the APR4 EOS, $\Ms=2.19M_{\odot}$, and $\Mc=1.39M_{\odot}$. The influences of $\Lambda$ and $\nu$ on $\sm$ will be postponed until Sec.~\ref{sec.V.B}.

The leading-order (5PN) tidal effects in GW waveforms, measured by the advanced Laser Interferometer Gravitational-wave Observatory (aLIGO) and other ground-based GW detectors, are encoded in the phase variation \citep{Flanagan08,Hinderer10,Favata14}
\begin{align}
	\Delta\varphi=-\frac{65}{4}\int \mathcal{M}^{-10/3} q_{\text{sym}} \Lambda\Oo^{2/3}d\Oo,
\end{align}
where the tidal deformability $\Lambda$ is given by
\begin{align}\label{eq:tidaldect}
	\Lambda \propto \frac{1+12q}{(1 + q)^5}\Lambda_1 + \frac{1+12/q}{(1 + 1/q)^5}\Lambda_2,
\end{align}
with $\Lambda_1$ and $\Lambda_2$ being the (dimensional) tidal Love numbers of the primary and the companion, respectively \citep{Hinderer08,Binnington09}. It has been shown that for $\mathcal{M}\lesssim1.5M_{\odot}$ and under the common EOS assumption, $\Lambda_1$ and $\Lambda_2$ relate to each other via [see Eq.~8 of \cite{De08}]
\begin{align}
	\Lambda_1 \simeq q^6 \Lambda_2,
\end{align}
translating Eq.~\eqref{eq:tidaldect} to
\begin{align}\label{eq:mutuallove}
	\Lambda(q,\Lambda_1) \propto \frac{12+q+q^2+12q^3}{q^2(1+q)^5} \Lambda_1.
\end{align}
As a result, $\Lambda$ decreases quite fast for large $q$, e.g., fixing $\Lambda_1$ and comparing a binary with $q=1.3$ (or $q_{\text{sym}}=0.246$) to an equal mass binary ($q_{\text{sym}}=0.25$), we find $\Lambda(1.3,\Lambda_1)/\Lambda(1,\Lambda_1)=0.47$. On that account, any GW-related constraints that might arise from the system are weaker. Further, the total emitted GW energy is a decreasing function of $q$ during both the inspiral and the post-merger phase \citep{Dietrich17}. In the ideal situation in the future where one observes a precursor and GWs from the same inspiral, unequal mass binaries provide marginally worse information from GWs even if they are more likely to cause crustal fractures. There is a trade off of sorts therefore.

\subsection{Dependence on Magnetic Field and Stratification} \label{sec.V.B}
According to the previous discussion, although crust failure tends to occur in a primary that is a member of a binary with small symmetric mass ratio, $\svom$ is insensitive to $q$ for a fixed $\Ms$ (see the discussion of Fig.~\ref{fig:chirp-q-wff1}). In addition, studying the whole multidimensional parameter space is laborious so that we concentrate on equal-mass binaries ($q=1$) in this subsection and emphasise the impact of magnetic field, which is parameterised by $\Bc$ and $\Lambda$, and stratification $\delta$ on $\sm$.

\begin{figure*}
	\subfigure{
		\includegraphics[scale=0.28]{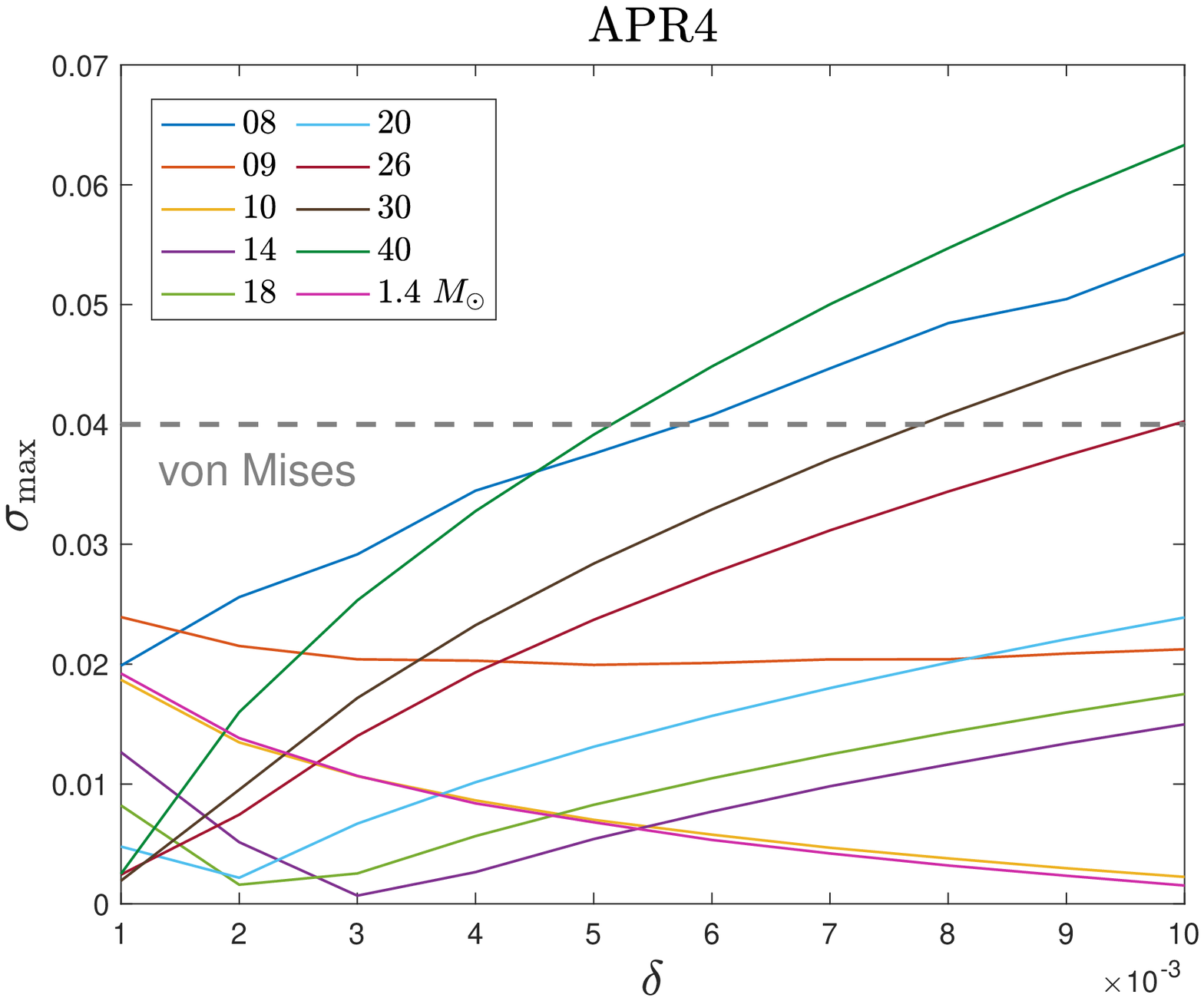} }
	\subfigure{
		\includegraphics[scale=0.28]{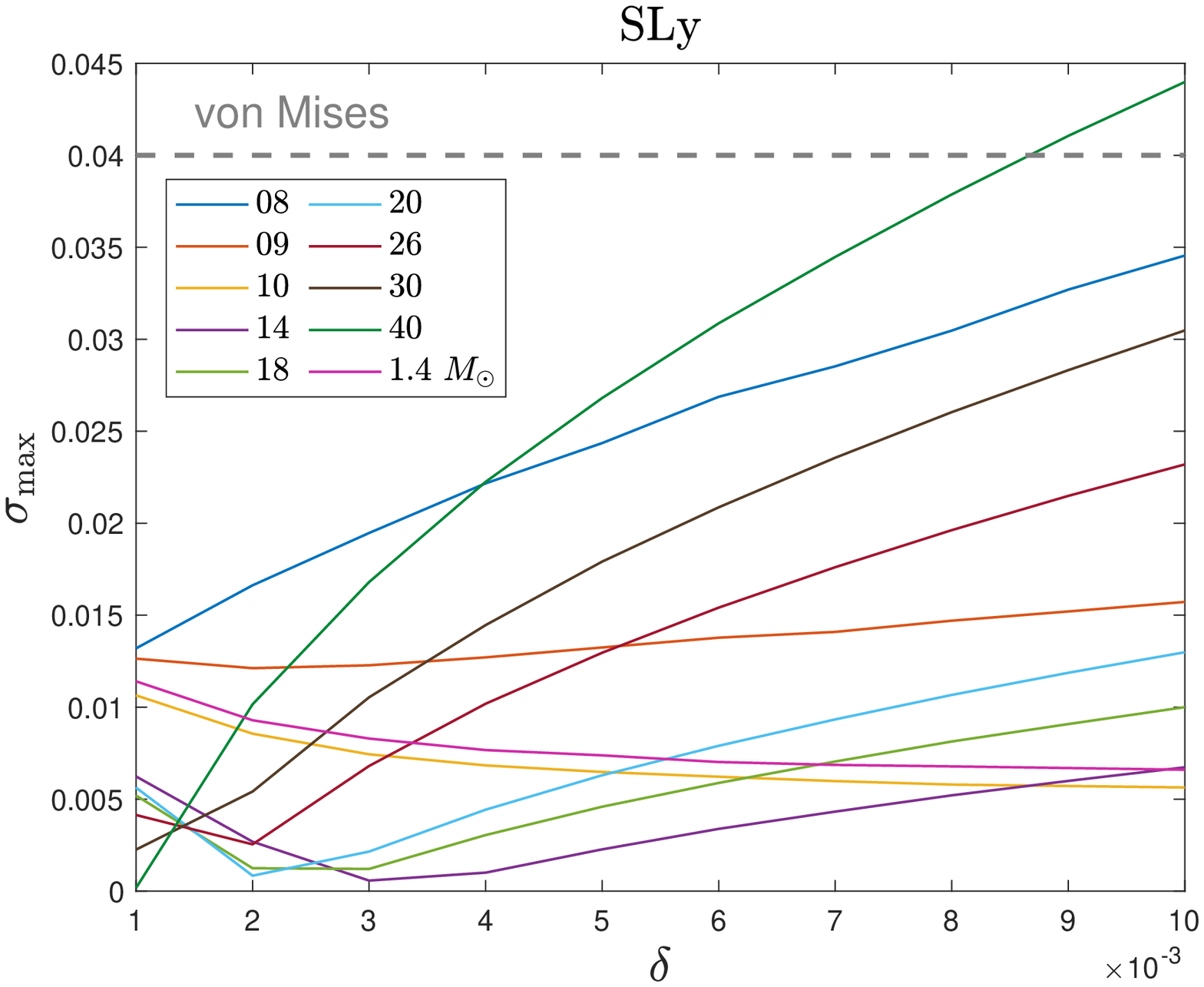}  }
	\subfigure{
		\includegraphics[scale=0.28]{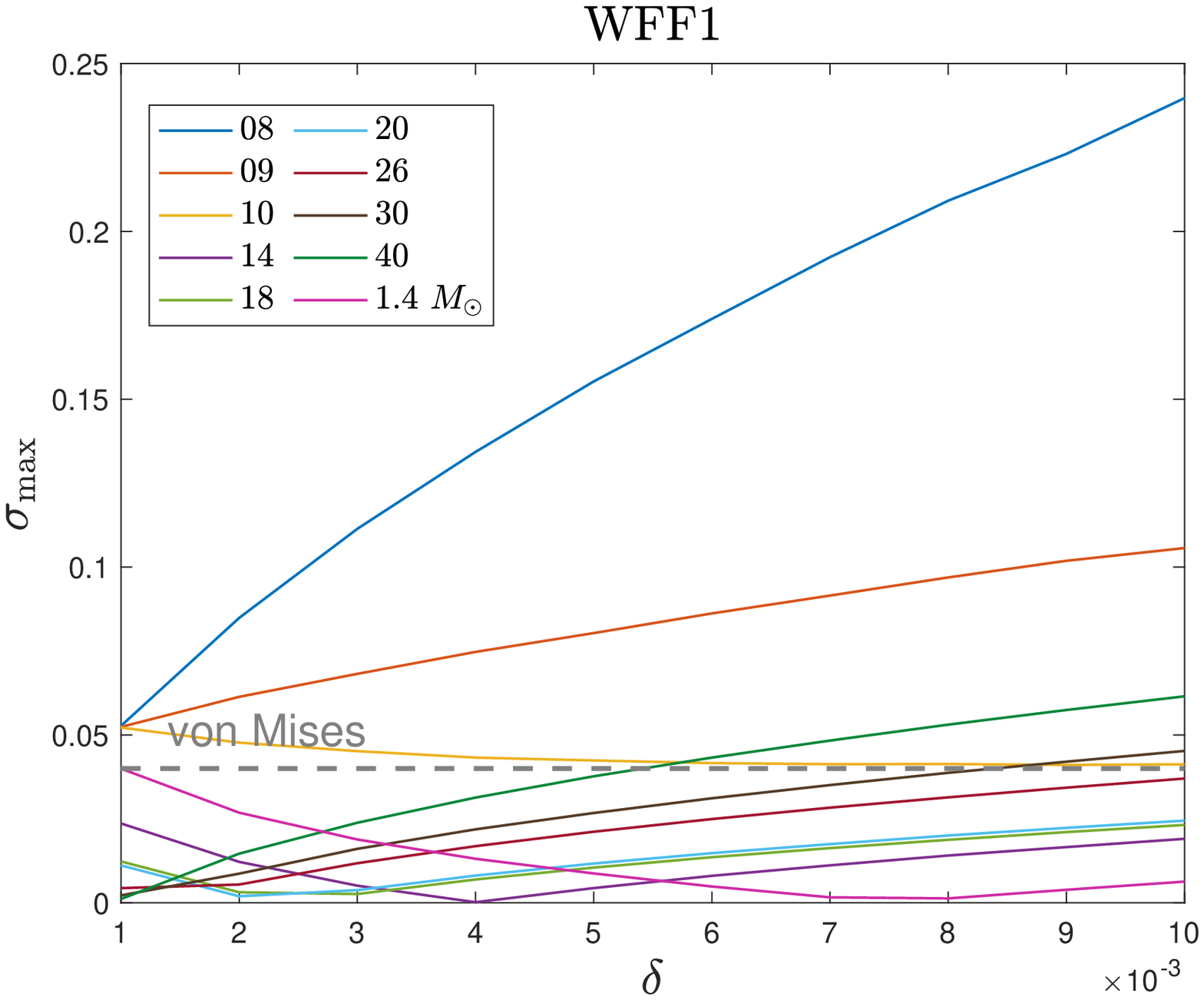}  }
	\subfigure{
		\includegraphics[scale=0.28]{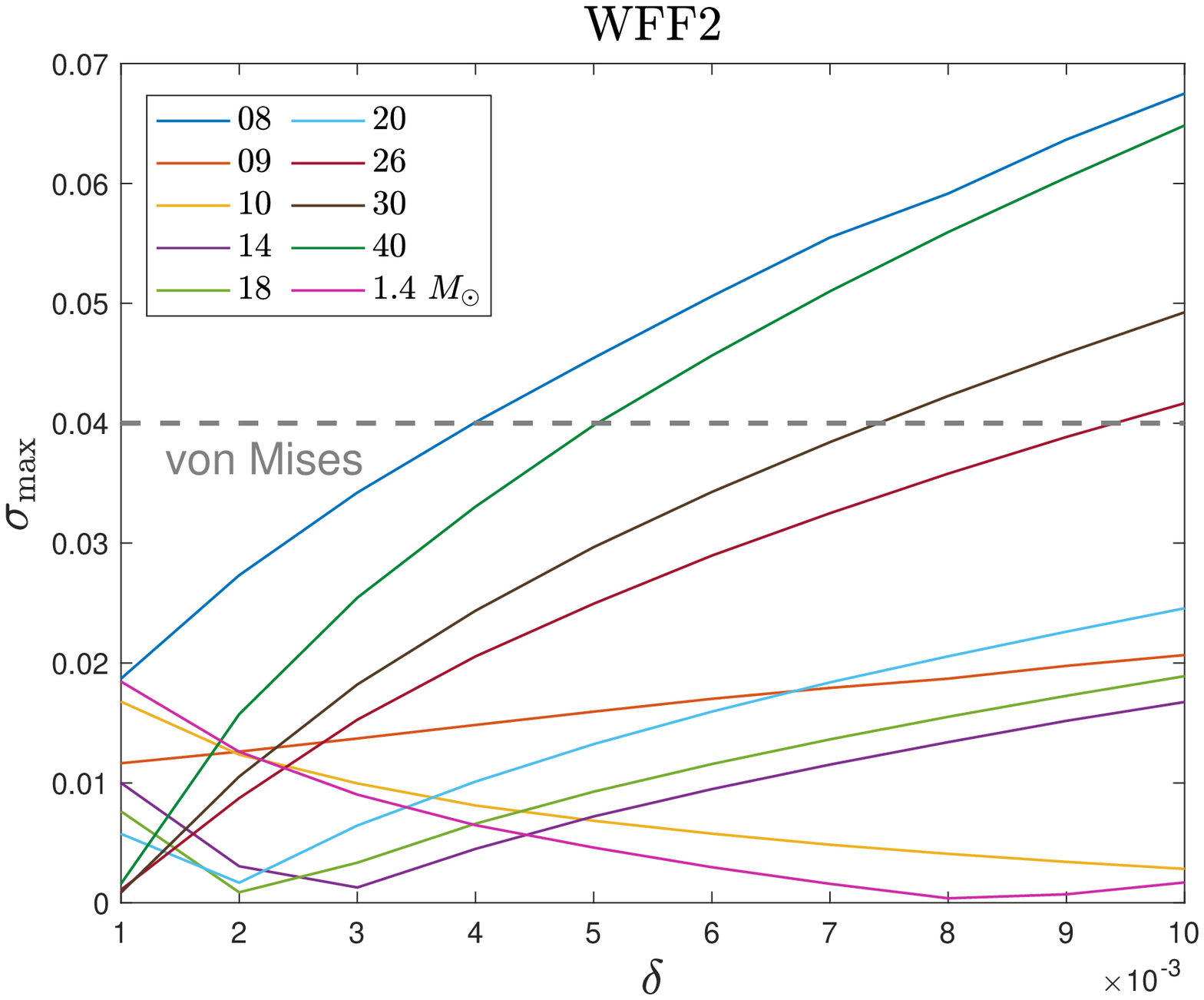}  }
	\subfigure{
		\includegraphics[scale=0.28]{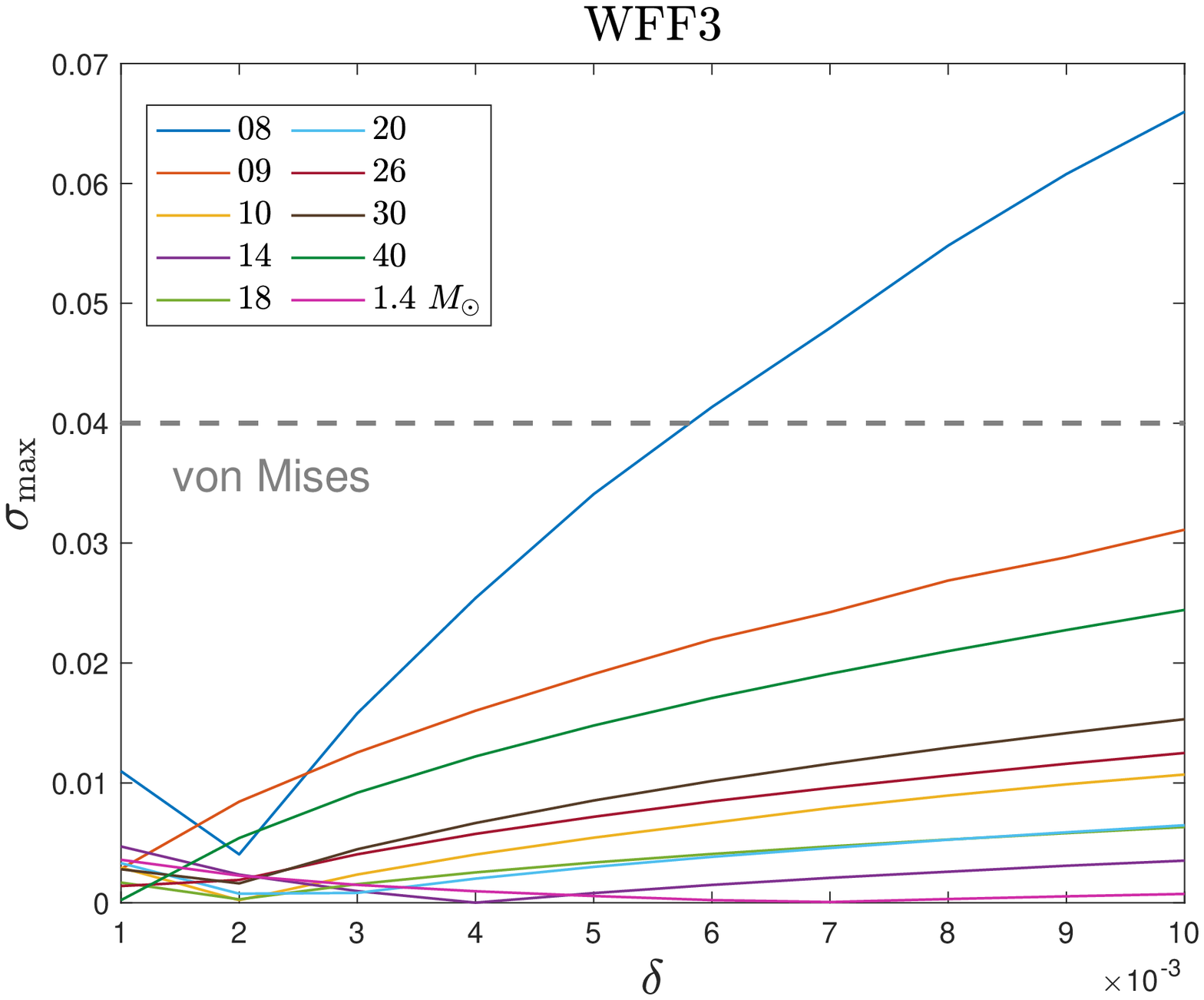}  }
	\caption{Maximal crustal strain $\sm$ due to $g_1$-modes of some chosen models of each EOS as functions of $\delta$. 
	The charateristic magnetic strength is fixed as $\Bc=2.5\times10^{15}$ G.
	The horizontal dashed lines mark the von Mises criterion.
	}
	\label{fig:maxstrain}
\end{figure*}

In Fig~\ref{fig:maxstrain} we show $\sm$ of $g_1$-modes for some models with the EOS introduced above as functions of $\delta$, where we set $\Lambda=1$ (purely poloidal) and $\Bc=2.5\times10^{15}$ G. 
Two kinds of tendencies are observed: (i) $\sm$ increases with $\delta$ for stars with either high or low compactness; (ii)  $\sm$ decreases with increasing $\delta$ for stars having moderate compactness. We see that $\sm$ for stars of the first tendency are larger, suggesting again the tidally-driven shattering favours stars with strong or weak gravity. In addition, the frequency of $g$-modes, as well as the tidal overlap, decreases with $\delta$; stipulating a small $\delta$, resonances happen at low orbital frequency thus have longer resonant duration (NSs shrink slowly at large separation), while the weaker coupling strengths limit the growth of mode amplitudes. One thus weights these two effects in to determine $\sm$, which can be roughly estimated by the product of resonant duration and tidal overlap. Inflection points exist on some curves for moderate-compact stars, where the resonant duration and the tidal overlap strength offset each other most. Right to these points, the large overlap compensates the short resonant duration, while the long resonance makes up the small tidal coupling for the other part. Additionally, we note that if we adopt $\svom\sim 0.1$, then all the cases presented in Fig.~\ref{fig:maxstrain} is not able to meet the von Mises criterion.

In line with recent suggestions by \cite{Passamonti:2020fur}, our calculations show that crustal failure can happen for a wider range of stellar models when $\delta$ is larger [cf.~Fig.~\ref{fig:maxstrain}], while the inclusion of magnetic fields and stellar rotation enriches the picture further. 
For both $g_{1}$- and $g_{2}$-modes, a pure poloidal magnetic field ($\Lambda=1$) shifts mode frequencies upward, leading to shorter resonant time. On the opposite, either the toroidal component of magnetic field or the rotation of the equilibrium configuration can give rise to a larger $\sm$ due to the negative shifts in mode frequencies. GRBs 090510 $a$ and $b$ have been accounted for by the rotation of the primary [Tab.~\ref{tab:results}]. In certain range of $\Lambda$, the mode modifications for $g_1$- and $g_2$-modes have different signs [see Fig.~(7) in \citetalias{Kuan21}].  The negative shifts for $g_1$-modes make them resonant with the earlier orbital frequency, while the positive shifts for $g_2$-modes delay their resonances. Therefore, $g_1$-modes will be resonantly excited \emph{prior to} $g_2$-modes if strong magnetic field is present.
For instance, setting $\Lambda=0.18$ and $\Bc=2.85 \times10^{15}\text{ G}$, we find that the resonances of $g_1$- and $g_2$-modes for the primary with EOS WFF1 and $\Ms=0.86M_{\odot}$ occur at, respectively, $t=12.95\text{ s}$ and $t=0.42\text{ s}$. Therefore, a magnetic field with toroidal component may present another scenario that may account for the two GRBs 090510a and 090510b.

\begin{figure}
	\subfigure{	\includegraphics[scale=0.44]{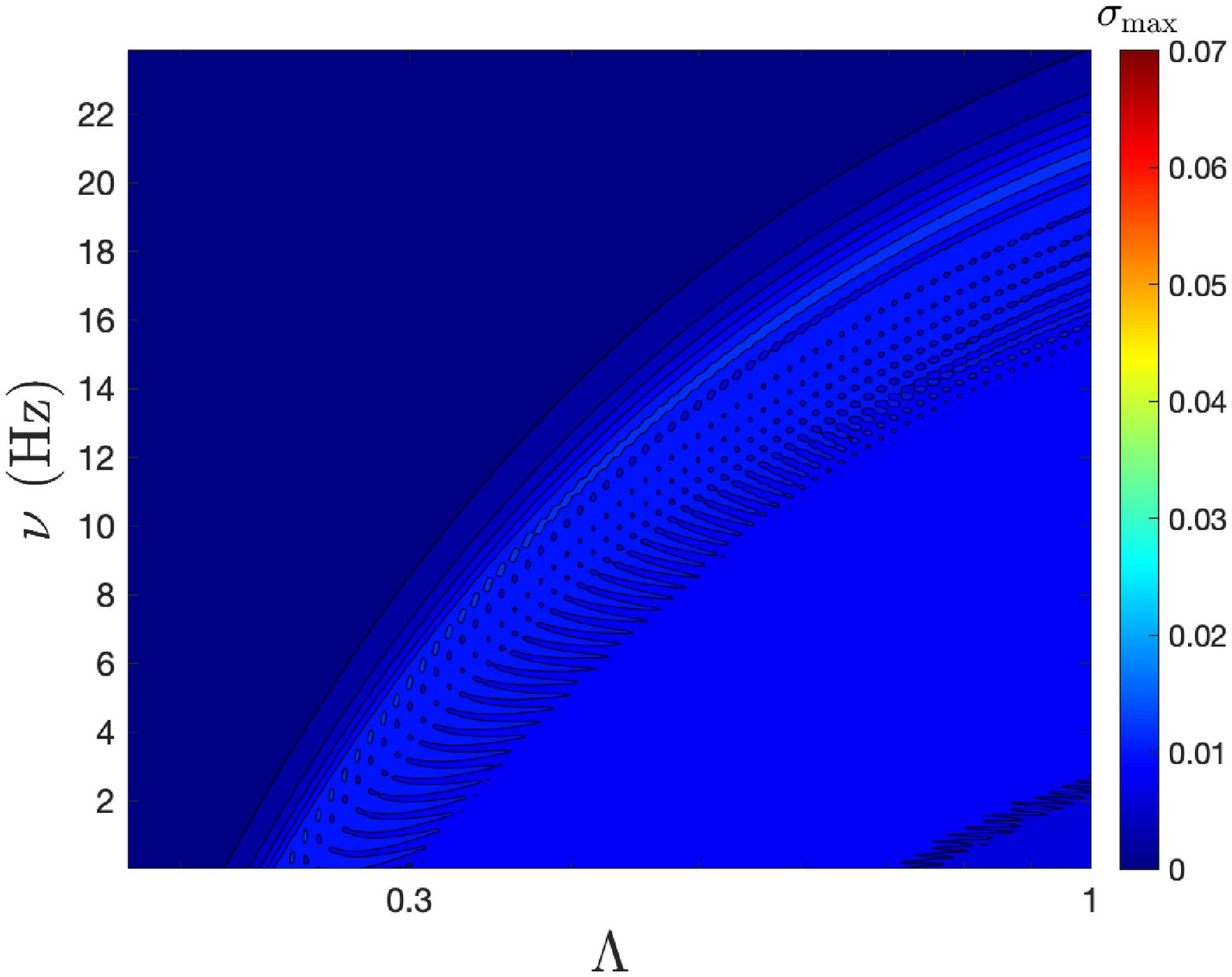} }
	\subfigure{	\includegraphics[scale=0.44]{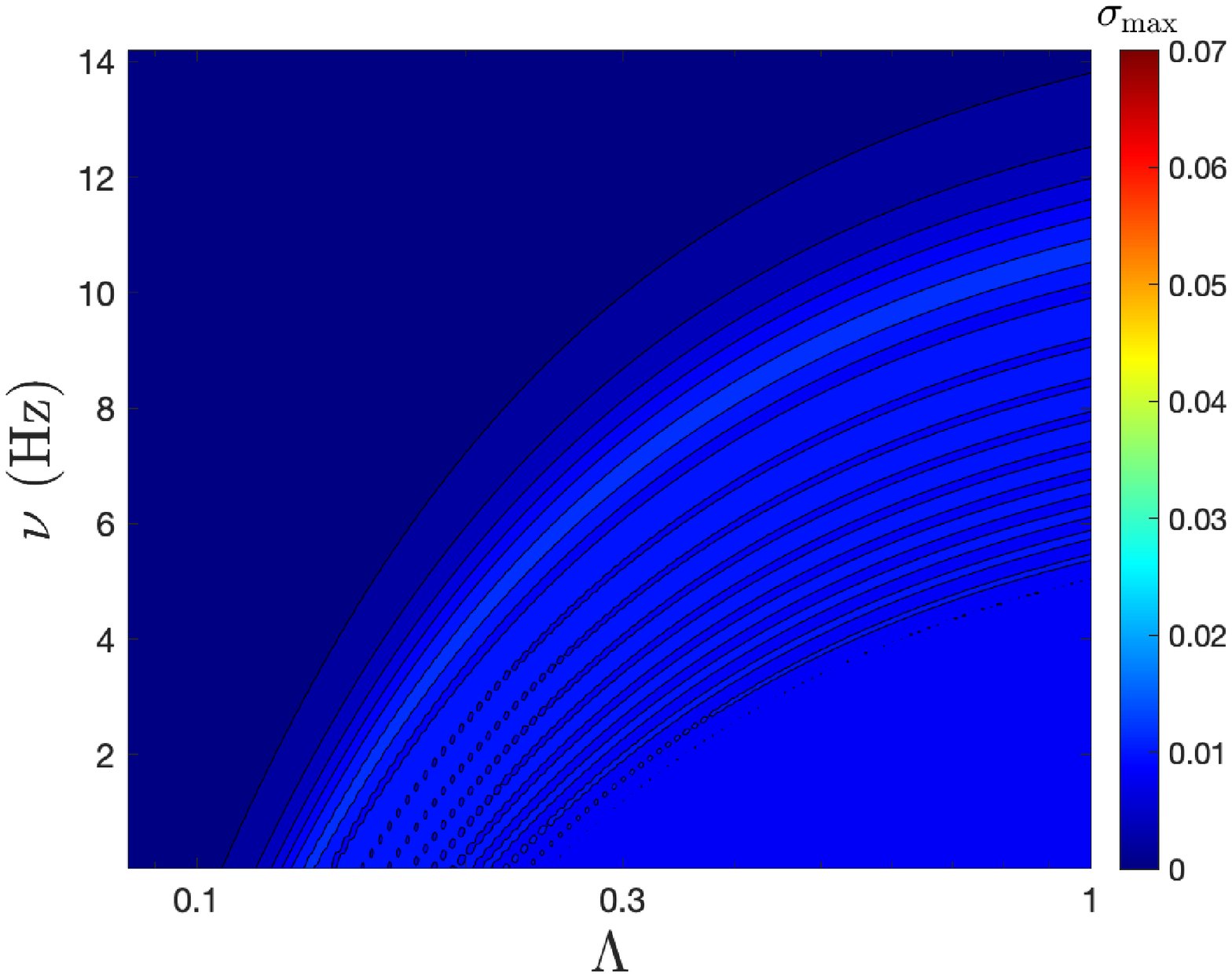} }
	\subfigure{	\includegraphics[scale=0.44]{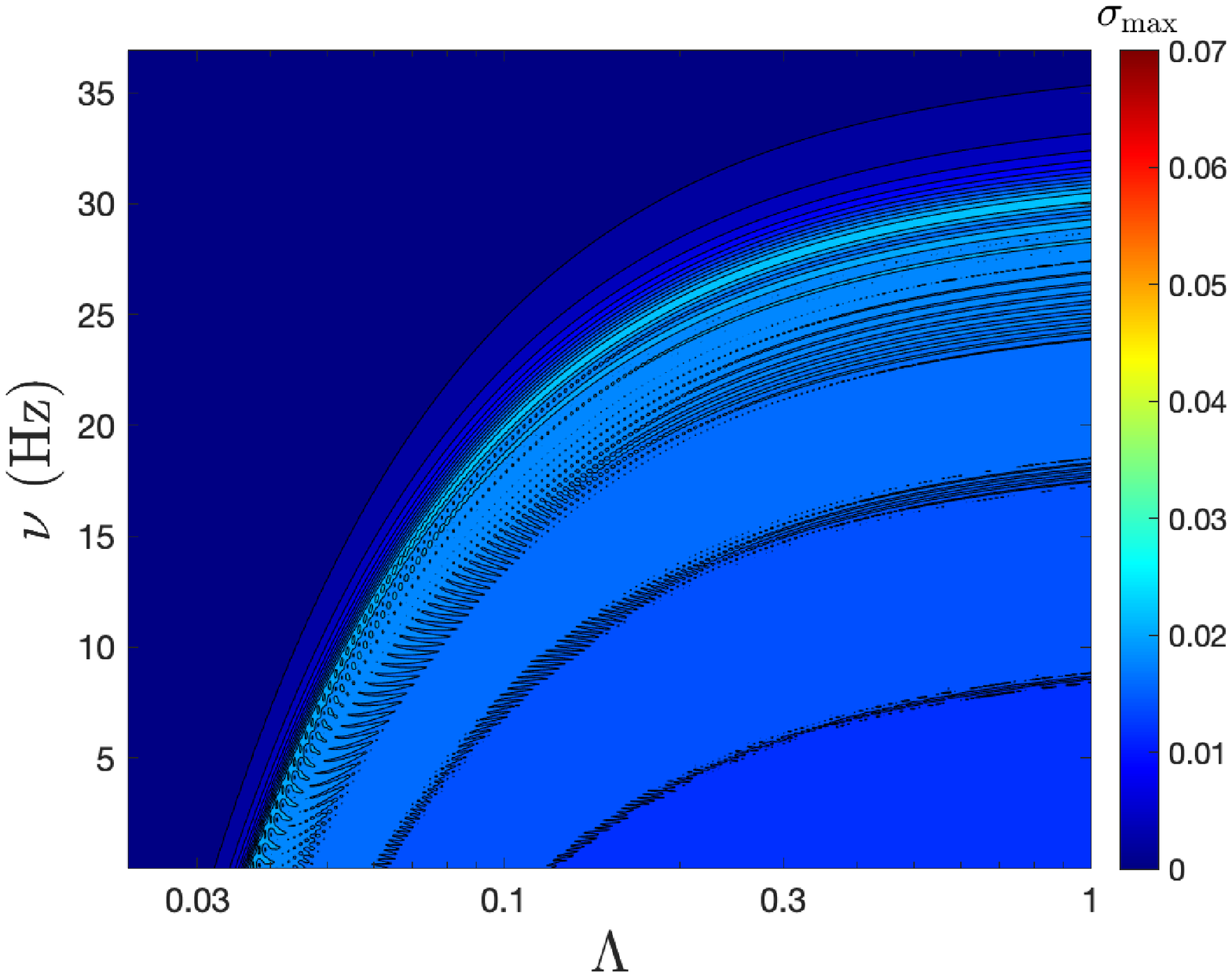} }
	\caption{Maximal crustal strain by $g_{1}$-modes as a function of $\delta$ and $\Lambda$ for the star with EOS SLy and $1.27 M_{\odot}$. Brighter shades indicate a greater value for $\sm$. The parameters $(\Bc,\delta)$ are, from top to bottom panel, taken as $(2,0.005)$, $(1,0.005)$, and $(1,0.01)$, where $\Bc$ is given in the unit of $\Bn$.
	}
	\label{fig:omega-lambda_g1}
\end{figure}

\begin{figure}
	\subfigure{	\includegraphics[scale=0.44]{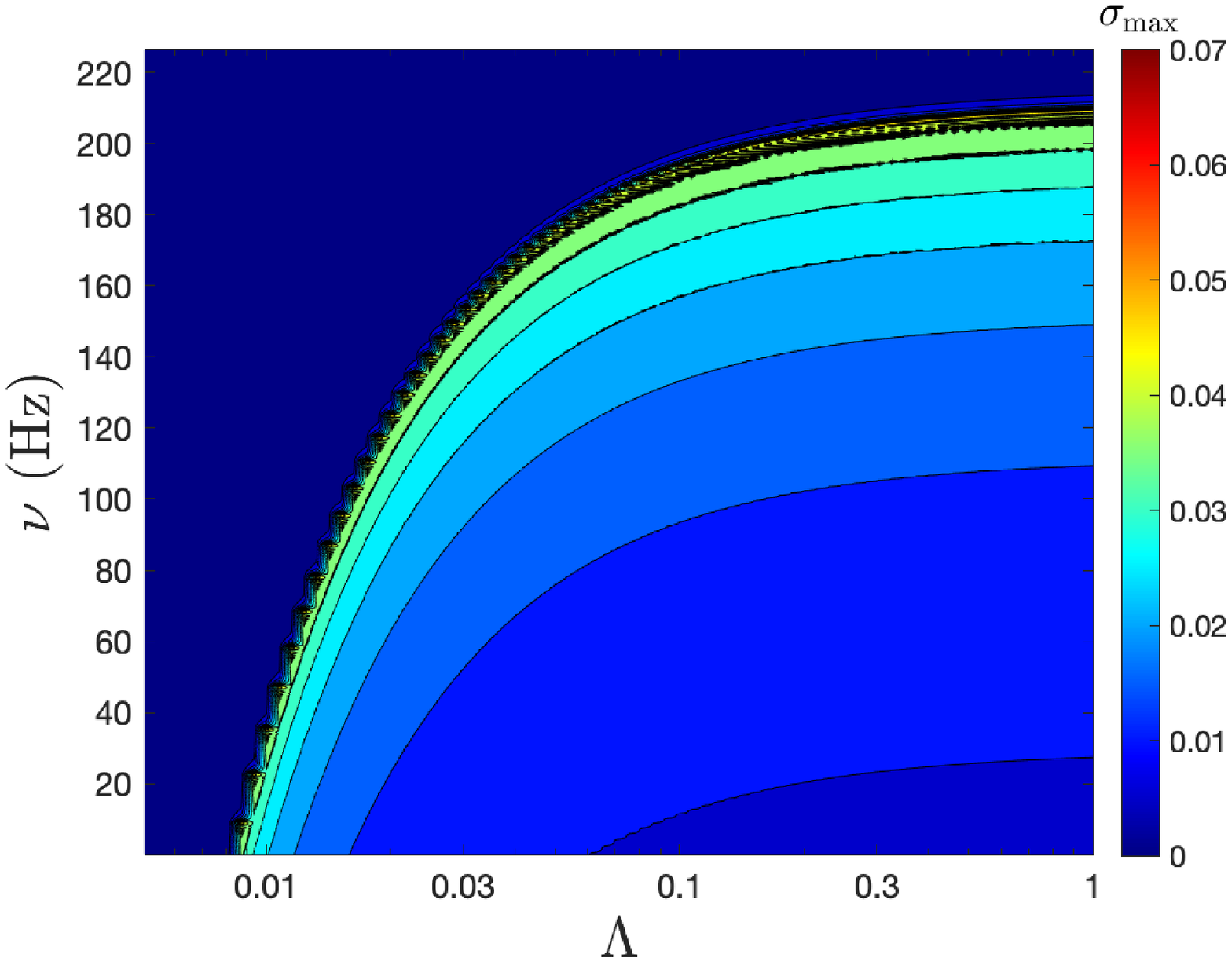} }
	\subfigure{	\includegraphics[scale=0.44]{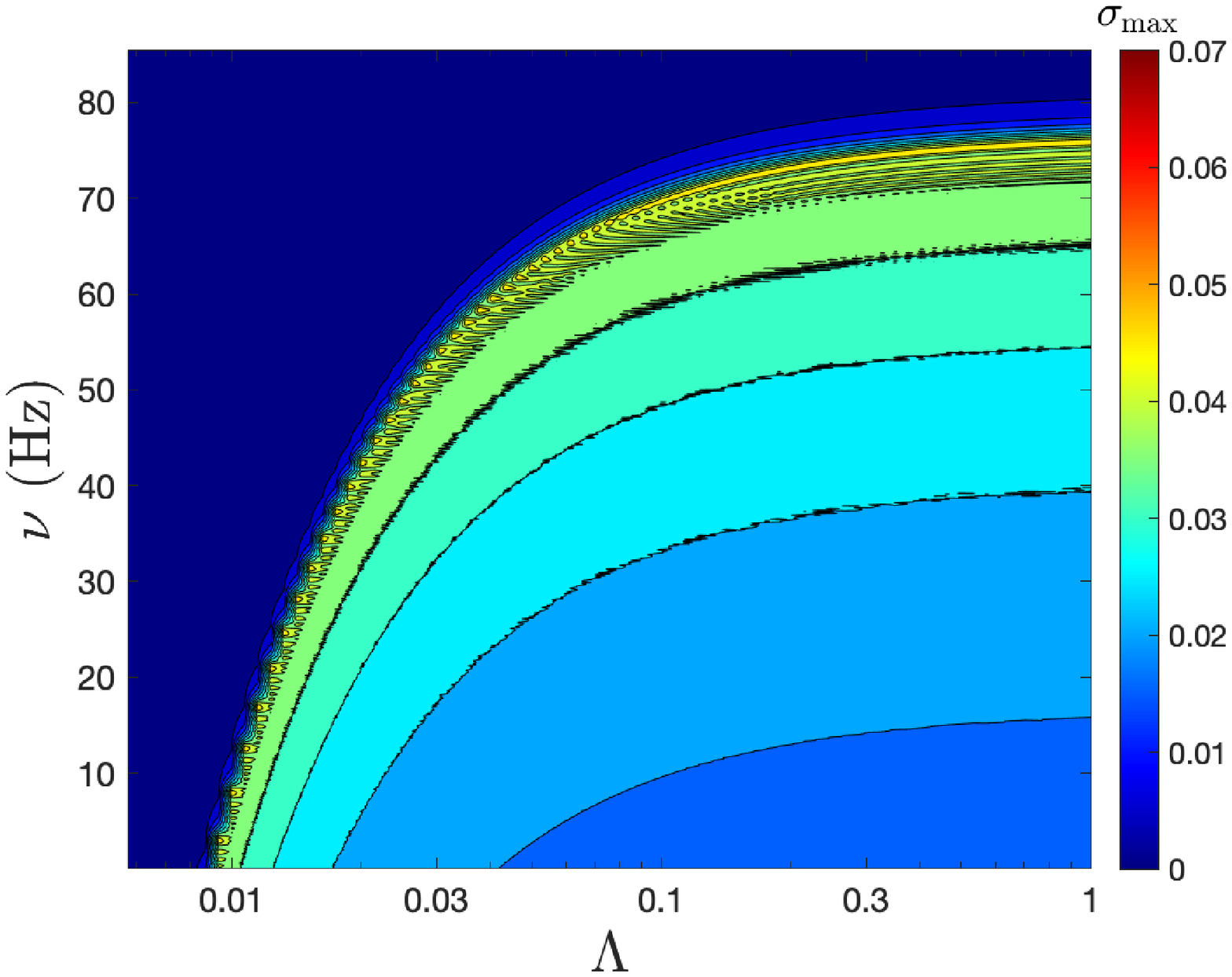} }
	\subfigure{	\includegraphics[scale=0.44]{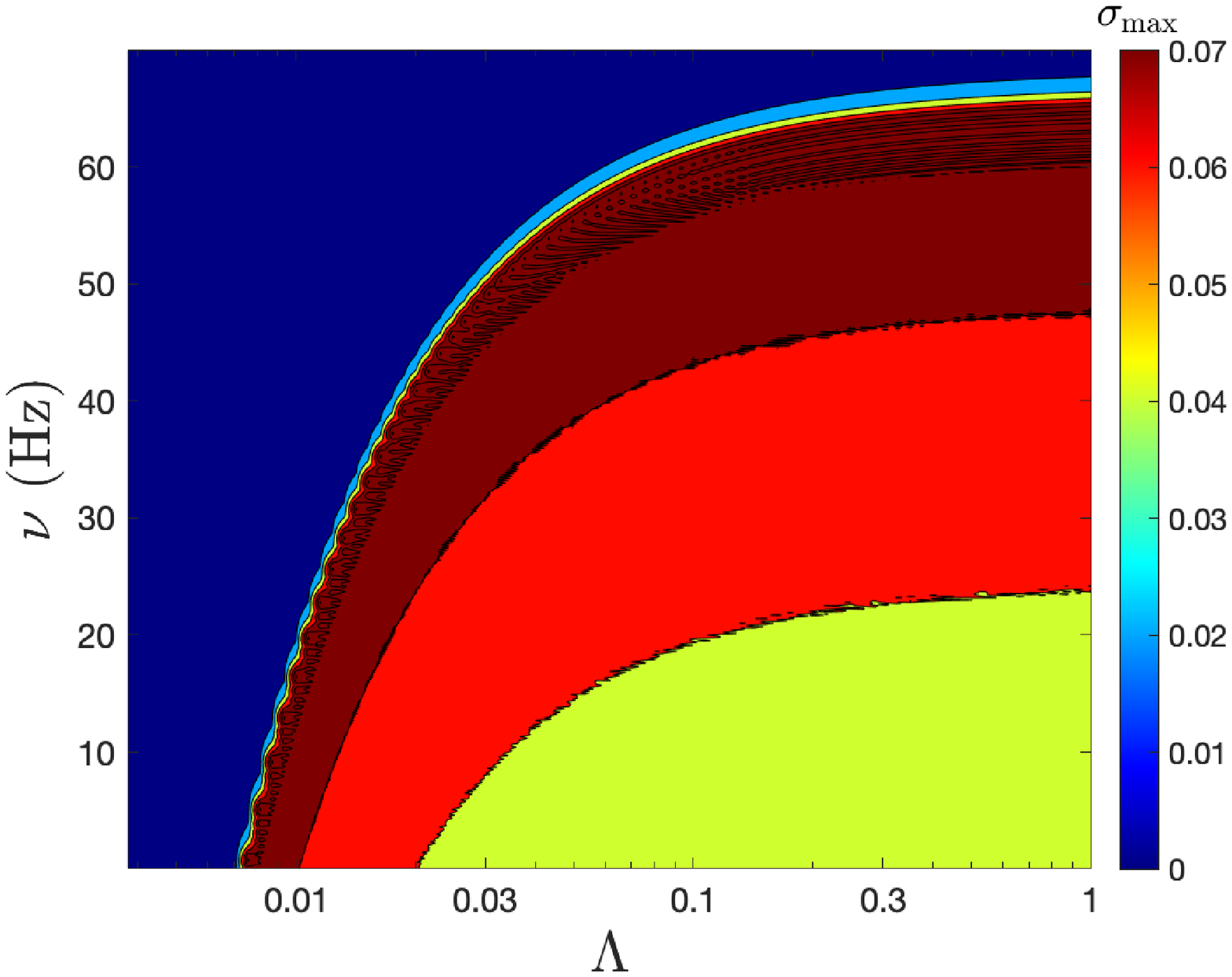} }
	\caption{Maximal crustal strain by $g_{2}$-modes as a function of $\delta$ and $\Lambda$ for the star with EOS SLy and $1.27 M_{\odot}$. Brighter shades indicate a greater value for $\sm$. The parameters $(\Bc,\delta)$ are, from top to bottom panel, taken as $(0.8,0.005)$, $(0.5,0.005)$, and $(0.5,0.01)$, where $\Bc$ is given in the unit of $\Bn$.
	}
	\label{fig:omega-lambda_g2}
\end{figure}

In Fig.~\ref{fig:omega-lambda_g1} and Fig.~\ref{fig:omega-lambda_g2}, we show $\sm$ by $g_{1}$- and $g_{2}$- modes for some fixed stellar parameters over the parameter space spanned by $\Lambda$ and $\nu$. It can be observed that $\sm$ for the $g_{2}$-mode reach values above $0.04$ for a certain region of the two dimensional parameter space, while the von Mises criterion is not met for the non-spinning model with pure poloidal magnetic field [top and middle panels in Fig.~\ref{fig:omega-lambda_g2}]. For both $g_{1}$- and $g_{2}$-modes, the optimal $\sm$ is two times higher than the non-rotating models with $\Lambda=1$.  Our results can be summerised as follows:
\begin{enumerate}
	\item When other parameters are fixed, the maximal crust strain $\sm$ is an increasing function of stratification $\delta$. Defining the optimal region as the set of combination of $\Lambda$ and $\nu$ for which $\sm$ is at the greatest level of the colorbar beside the figures, a  lower $\Lambda$ (stronger toroidal field) or higher spin is necessary for the optimal case. In addition, the optimized $\sm$ does not depend on $\Bc$.
	\item Although $\Bc$ changes the pattern of $\sm$ as a function of $\Lambda$ and $\Omega$, the value of $\sm$ remains unchanged. In addition, the optimal situation for $g_{1}$- and $g_2$-modes with stronger $\Bc$ requires faster spins. 
	\item Over the optimal region of each mode, $g_{2}$-modes cause stronger strains $\sm$ than $g_{1}$-modes. Yet, optimal cases of $g_2$-modes require the magnetic field to have a dominant toroidal field ($\Lambda\approx10^{-2}$), which thus constrains the maximum alowed values of $\Bc$ \citep{Reisenegger:2008yk}. 
\end{enumerate}
Although the effects of rotation and magnetic field create some room for potential crust failure in the parameter space, the parameters should still be fine-tuned to generate a strain $\sm>\svom$. In other words, a given precursor event may set stringent constraints on the properties of the individual stars in a binary.

\section{Discussion}\label{sec.VI}

With multi-staged SGRBs, namely those including precursors, main events, and afterglows, we can garner better knowledge about the properties of the progenitors and fundamental physics governing NSs, such as their EOS.  For instance, strongly magnetized remnants from binary mergers, as inferred from X-ray plateaus observed in some SGRB afterglow light curves \citep{Rowlinson13,Gompertz:2013zga} or early X-ray flares observed in SGRB light curves \citep{Gao:2005yd}, may hint that the progenitors consist of at least one highly magnetized NS from a flux conservation argument \citep{Ciolfi19}. Detailed studies of SGRBs may also unveil the nature of their central engines. Analysing precursors may therefore shed light on the qualitative properties of the progenitors, and could tightly constrain the stellar parameters of the merging stars (\citealt{Tsang:2011ad,Tsang:2013mca,Passamonti:2020fur}; \citetalias{pap1}; \citealt{Neill21}). 

To explore the connection between crustal fractures and precursors, we adopt the theoretical framework detailed in \citetalias{Kuan21}. To briefly recall, we consider the tidal resonance between QNMs and the orbit, where we treat general-relativistic QNM spectra, and the orbital dynamics involves up to the 3 PN effects including the 2.5 PN scheme for gravitational back-reaction. The modification of mode frequencies by perturbing forces from magnetic fields [Eq.~\eqref{eq:modmag}], tidal field [Eq.~\eqref{eq:modtid}], and stellar rotation [Eq.~\eqref{eq:modrot}] are also taken into account.
When a particular mode is brought into resonance -- as defined by the time interval when the orbital frequency and the (modified) mode frequency are matched to some extent [Eq.~\eqref{eq:resnbh}] -- the mode amplitude increases rapidly. If the maximal amplitude available during a resonant timescale, the crustal fracture may be caused. Over the yielding area, stored energy will be released in some form [Eq.~\eqref{eq:availerg}].
Taking a particular binary and some fixed stellar parameters, we match the data of precursors by varying $\Bc$ to make the onset of resonances coincide with the moment (relative to the main event) precursors are detected (Tab.~\ref{tab:results}). Assuming all released energy is transformed into electromagnetic radiation [estimated in the 6th column of Tab.~\ref{tab:results}], SGRB precursor events may be accommodated, energetically speaking, by crust failure. On the other hand, we present two scenarios for SGRBs hosting two precursors, e.g.~for GRB 090510 either the spin down of the primary leads to the same mode gets resonant twice at different moments [Tab.~\ref{tab:results}] or the resonances of $g_1$- and $g_2$-modes when the magnetic field has toroidal component.

We find that for a given primary, a relatively large mass ratio is more favourable for crustal fracture (Fig.~\ref{fig:chirp-q-wff1}). However, the price to pay is the detectability of the tidal imprints in GW [Eq.~\eqref{eq:mutuallove}]. In addition, we find that for certain combination of stellar parameters the von Mises criterion can be met. For instance, when $\nu$ and $\Lambda$ are tuned to particular values [the brightest region in Fig.~\ref{fig:omega-lambda_g1} and Fig.~\ref{fig:omega-lambda_g2}] the strain exceeds $\svom$. In other words, as long as precursors prove to set constraints on the properties of progenitors, the constraints are going to be stringent because several parameters are limited simultaneously.

Tidal effects have been studied in various aspects, such as from the GW energy spectrum \citep{Faber02,Bauswein19}, the NS tidal disruption signal for binaries having a least one NS \citep{Vallisneri00,Ferrari10}, and numerical simulations of NSNS mergers [see \cite{Baumgarte03} for a detailed review]. The aforementioned investigations are devoted to the prospect of extracting information about the EOS from the very final stage of inspiral [$f_{\text{GW}}\sim 1000\text{ Hz}$ \citep{Kokkotas05}]. Precursors, however, offer an extra probe into the details of EOS when $f_{\text{GW}}\sim 100\text{ Hz}$ [cf.~Fig.~\ref{fig:SGRBs}], which is also the most sensitive band of (ground-based) interferometers such as aLIGO, Virgo, and KAGRA \citep{Moore15,Schmitz21}. Therefore, an application of this framework to future precursor data together with prospective GW detections may result in strong tests of the neutron star EOS \citep{Zink12}.

One major limitation of our approach is the absence of a solid crust. The imposition of an elastic crust introduces a localised, non-zero shear modulus $\mu$ to the outer most $\sim 1$ km of the star, which leads to a shear stress tensor quenching perturbations there. In particular, the rigid crust attenuates fluid motions near the crust-core interface due to the discontinuity in $\mu$, rendering slippages in the (tangential) displacements [see, e.g., \cite{Kruger15,Passamonti:2020fur}]. The degree of the slippage can be viewed as the extent to which the restoring forces of modes are balanced by the elasticity, and therefore depends on the shear modulus, and the thickness of the crust. The slippage naturally tends to zero in the crustless limit, i.e., when $\mu\rightarrow 0$.

When some specific restoring force dominates over the elasticity, the crust will become susceptible to the corresponding fluid motion. As has been investigated in the literature, (i) the restoring forces for $p$- and $f$-modes tend to overwhelm the elasticity, so that the presence of the crust impacts only slightly their eigenfunctions and frequencies \citep{McDermott88}, (ii) \cite{Levin01} found that leading order $r$-modes will strongly couple to the crust in the sense that the associated eigenfunctions are not damped at the boundary interface (i.e., the slippage is quite small) when the star spins fast enough such that the centrifugal force eclipses the rigidity, and (iii) \cite{Colaiuda11} demonstrated the eigenfunction of Alfv\'en modes will extend into the crust when the magnetic field is sufficiently strong. These results suggest that there exists a threshold on the mode frequency above which the eigenfunction may penetrate the crust to some extent. As it requires a more in depth discussion [see, e.g., \cite{Glampedakis06} for the analysis of $r$-modes], we will address the problem for $g$-modes elsewhere, thereby re-examining the relevance of them as triggers for precursors.

\section*{Acknowledgement}
This work was supported by the Alexander von Humboldt Foundation, the Sandwich grant (JYP) No.~109-2927-I-007-503 by DAAD and MOST, and the DFG research Grant No. 413873357. We thank David Tsang for providing helpful feedback, which improved the quality of the manuscript.

\section*{Data availability statement}
Observational data used in this paper are quoted from the cited works. Data generated from computations are reported in the body of the paper. Additional data can be made available upon reasonable request.

\label{lastpage}

\end{document}